\documentclass[preprint,12pt]{elsarticle}

\usepackage{amssymb}

\usepackage{amsmath}

\usepackage{color}

\newcommand{\delf}{\delta f} 
\newcommand{\dld}[2]{\frac{\partial{#1}}{\partial{#2}}}
\newcommand{\dldl}[2]{\partial{#1}/\partial{#2}}
\newcommand{\dd}[2]{\frac{d#1}{d#2}}
\newcommand{\vpr}{v_\parallel}
\newcommand{\vpp}{v_\perp}
\newcommand{\vv}{\mathbf{v}}
\newcommand{\xx}{\mathbf{x}}
\newcommand{\fave}[1]{\left\langle{#1}\right\rangle}
\newcommand{\ddct}{\left.\dd{}{t}\right|_{C^{T0}_{ab}}}

\journal{Computer Physics Communications}

\begin{document}

\begin{frontmatter}


\author[nifs,soken]{Shinsuke Satake\corref{cor1}}
\ead{satake@nifs.ac.jp}

\author[nifs,soken]{Motoki Nataka}
\author[vist]{Theerasarn Pianpanit}
\author[nifs,soken]{Hideo Sugama}
\author[nifs,soken]{Masanori Nunami}
\author[nifs,soken]{Seikichi Matsuoka}
\author[nifs,soken]{Seiji Ishiguro}
\author[nifs,soken]{Ryutaro Kanno}

\cortext[cor1]{Corresponding author}
\address[nifs]{National Institute for Fusion Science, Toki 509-5292, Japan}
\address[soken]{The Graduate University for Advanced Studies, Toki 509-5292, Japan}
\address[vist]{School of Information Science and Technology, Vidyasirimedhi Institute of Science and Technology, Rayong, Thailand}

\title{Benchmark of a new multi-ion-species collision operator for $\delta f$ Monte Carlo neoclassical simulation}


\begin{abstract}
A numerical method to implement a linearized Coulomb collision operator in the two-weight $\delta f$
Monte Carlo method for multi-ion-species neoclassical transport simulation is developed.
The conservation properties and the self-adjoint property of the operator in the collisions between two particle 
species with different temperatures are verified. The linearized operator in a $\delta f$ Monte Carlo code
is benchmarked with other two kinetic simulations, a $\delta f$ continuum gyrokinetic code
with the same linearized collision operator and a full-f PIC code with Nanbu collision operator. The benchmark 
simulations of the equilibration process of plasma flow and temperature fluctuation among several particle 
species show very good agreement between $\delta f$ Monte Carlo code and the other two codes. 
An error in the H-theorem in the two-weight $\delf$ 
Monte Carlo method is found, which is caused by the weight spreading phenomenon inherent in 
the two-weight $\delf$ method. It is demonstrated that the weight averaging method serves to restoring 
the H-theorem without causing side effect.
\end{abstract}


\end{frontmatter}


\section{Introduction}\label{sec1}
To study and evaluate the neoclassical transport\cite{neo1,neo2,neo3}
in burning plasmas such as in ITER\cite{iter} 
and future fusion reactors, it is essential to consider the 
transport process in a plasma containing deuterium, tritium, helium from D-T fusion reaction, and 
impurity ions such as C, Fe, and W from the vacuum vessel and divertor wall. 
Neoclassical transport, which is described by drift-kinetic equation for guiding-center distribution function, 
treats the transport process of the charged ions and electrons in toroidal magnetic field caused by guiding-center drift motion and 
Coulomb collisions. Therefore,
it is required to develop a neoclassical transport simulation code for multi-ion-species plasmas 
which treats the Coulomb collisions among unlike ion species and electrons. One of the important points
in treating the transport process in fusion plasma is that the large differences in the masses of the charged particles. 
Not only between ions and electrons of which mass ratio is $m_i/m_e> 10^3$, the mass ratio between 
bulk fuel D or T and heavy impurity ions like W can also be order of $10^2$. 
Basically, thermal equilibration between two particle species with large mass ratio is a slow process
compared to the momentum transfer among them.
Temperature difference among ion species has been paid little attention in the transport analysis so far,
mainly because reliable measurement of bulk ion temperature has become available only 
recently.\cite{tiexp1} Finite temperature difference among ion species is observed
in multi-species plasmas when strong ion heating is applied or at the pedestal region of
H-mode plasma, for example\cite{tiexp2,tiexp3}.

Most of numerical approaches for neoclassical transport simulation have been developed by 
using the linearized Fokker-Planck collision operator, or so called Landau operator, as it is explained in
Section 2.  The linearized collision operator\cite{hinton} has important properties for thermodynamics, 
that is, the self-adjointness of the operator and the Boltzmann's H-theorem. The former is related to
Onsager symmetry of the neoclassical transport matrix\cite{sgm-hort}, and the latter 
is understood as the non-negative nature of the entropy production rate by collisional process. 
It represents the dissipative nature of collisional transport process, which damps the distribution function 
toward a local equilibrium state. On the other hand,
Onsager symmetry appears as a consequence of the time reversibility of the underlying microscopic dynamics, 
which is the charged particle motion under the Coulomb interaction in the present case.
Also, in the application of the linearized collision operator in the drift-kinetic equation, it is shown that the 
entropy production rate coincides with the sum of the inner products of the thermodynamic forces 
and the conjugate neoclassical fluxes\cite{sgm-hort}. Therefore, the self-adjoint property and 
the H-theorem of the linearized operator make the basis of thermodynamics aspects of neoclassical transport theory.

However, it is known that  these two properties above are not rigorously satisfied by the Landau operator
when the particle species have different temperatures.\cite{hinton, sgm-c} 
Therefore, in many of previous studies on the neoclassical transport in multi-ion-species plasmas, 
it has usually been assumed that all the ion species have the same temperature. Only the temperature 
difference between ions and electrons are allowed, since the coupling of ions and electrons in the drift-kinetic
equation by the collision term is usually unimportant because of its large mass-ratio.  
As mentioned above, however, two ion particle species with large mass-ratio are also possible to have 
different temperatures. Therefore, to simulate the transport process in fusion reactors, it is advantageous to develop a collision operator for unlike-species collisions among the ion species with different temperatures. 
Sugama {\it et al.}\cite{sgm-c} has proposed a modified Landau operator 
which keeps the self-adjoint nature even if the temperature of two ion species are different.
We plan to extend the neoclassical simulation code FORTEC-3D\cite{f3d1,f3d2} by implementing the 
modified linearized Coulomb collision operator for multi-ion-species plasmas.
In this paper, a numerical method to implement the collision operator in a $\delta f$
Monte Carlo particle code like FORTEC-3D is explained. Benchmark of the new collision operator with 
the same one implemented in a continuum gyrokinetic $\delta f$ code GKV\cite{nnm, nkt} and the 
other one in a full-f PIC code\cite{pianpic} based on Nanbu-method\cite{nmb} are carried out. 
It will be demonstrated that the 
linearized operator implemented in the $\delf$ Monte Carlo code agrees well with the other two 
codes and also keeps the conservation and self-adjoint properties with high accuracy. 

The rest of this paper is organized as follows. In Section 2, basic properties of the linearized collision operator
are reviewed. The numerical method to implement the Sugama's modified Landau operator to the $\delf$ Monte
Carlo method is explained in Section 3. Benchmark tests of the collision operator with two different simulation 
codes are shown in Section 4, where we also discuss the numerical error appearing in the H-theorem 
by the weight-spreading effect and the way how to suppress it. Finally, the summary is given in Section 5.
In \ref{apdx_a} and \ref{apdx_b}, the details of the Monte Carlo scheme to implement the test-particle operator is explained. \ref{apdx_c} shows the simplified form of the linearized collision operator in the case two particle species
have the same temperature.
In this paper, we concentrate on the development of the collisions among ion species, and the numerical problem 
in the application of the collision operator for electron-ion collisions is discussed in Section 5 and \ref{apdx_a}.

\section{Properties of linearized collision operator}\label{sec2}
In a $\delta f$ drift-kinetic simulation code such as FORTEC-3D, the time evolution of a plasma distribution function 
of a particle species $a$ caused by Coulomb collisions and guiding-center motion in toroidal magnetic field is 
represented by the following drift-kinetic equation with the linearized Landau collision operator $C^L_a$, 
\begin{equation}
\dld{\delf_a}{t}+\dot{\mathbf{Z}}\cdot\dld{}{\mathbf{Z}}\delf_a=-\dot{\mathbf{Z}}\cdot\dld{}{\mathbf{Z}}
f_{Ma}+C_a^L(\delf_a),\label{dke}
\end{equation}
where $\mathbf{Z}=(\xx,\vpr,\vpp)$ is the 5-dimensional phase space coordinates. $\xx$,
$\vpr$, and $\vpp$ are the guiding-center position, particle velocity parallel and perpendicular 
to magnetic field, respectively.
$\delf_a(\mathbf{Z},t)=f_a-f_{Ma}$ represents the perturbation part of the distribution function from the
Maxwellian $f_{Ma}=n_a\left(\frac{m_a}{2\pi T_a}\right)^{3/2}\exp\left(-\frac{m_av^2}{2T_a}\right)$, 
where $n_a$, $m_a$, and $T_a$ are density, particle mass, and temperature of the particle species $a$, respectively, and $v=\sqrt{\vpr^2+\vpp^2}$ represents the particle absolute velocity.
The linearized collision operator $C^L$ is composed of the test- and field-particle operators, $C^T$ and $C^F$, respectively,
for all combinations of colliding particle species $a$ and $b$: $C_a^L(\delf_a)=\sum_b \left[C^T_{ab}(\delf_a,f_{Mb})+
C^F_{ab}(f_{Ma},\delf_b)\right]$. In the following, we use the abbreviations $C^T_{ab}(\delf_a,f_{Mb})=C^T_{ab}(\delf_a)$ and
$C^F_{ab}(f_{Ma},\delf_b)=C^F_{ab}(\delf_b)$.
Note that the summation over the colliding particle species $\sum_b$ includes the self-collision, $b=a$.  
The nonlinear term $C_{ab}(\delf_a,\delf_b)$ is omitted by the ordering assumption $|\delf/f_{M}|\ll 1$.
Distribution function is averaged over the fast gyration motion around the magnetic field, so the 
velocity coordinates are truncated into 2-dimensions, $(\vpr,\vpp)$. 
Note that throughout this paper, we discuss only on the time evolution of distribution function solely
by Coulomb collisions in a uniform plasma. Therefore,  instead of Eq. (\ref{dke}), we consider 
\begin{equation}
\dd{}{t}\delf_a(\vpr,\vpp,t)=C^L_a(\delf_a).\label{dkemod}
\end{equation}
In the numerical benchmark, the plasma is assumed to be in a uniform magnetic field, and 
the magnetic field strength satisfies the condition that Landau collision operator assumes, $\lambda_D/\rho_L\ll 1$, 
where $\lambda_D$ is the Debye length and $\rho_L=mv/eB$ is the Larmor radius. In the application 
of the linearized collision operator, it is also assumed the gradient scale length of background $n$ 
and $T$, $L_H=|\nabla \ln(n, T)|^{-1}$, is longer than the Debye length, $\lambda_D/L_H\ll 1$.

The linearized collision operator should satisfy the following properties:
\begin{subequations}
\begin{align}
\int d^3v C_{ab}^T(\delf_a)&=\int d^3v C_{ab}^F(\delf_b)=0,\label{cnsvn} \\
m_a\int d^3v \{\vv, v^2\} C_{ab}^T(\delf_a)&
=-m_b \int d^3v \{\vv, v^2\}C_{ba}^F (\delf_a), \label{cnsvpe}\\
\int d^3v \frac{\delta f_a}{f_{aM}} C_{ab}^T(\delta g)&=
\int d^3v \frac{\delta g}{f_{aM}} C_{ab}^T(\delta f_a),\label{eqadjt}\\
T_a\int d^3v \frac{\delta f_a}{f_{aM}} C_{ab}^F(\delta f_b)&=
T_b\int d^3v \frac{\delta f_b}{f_{bM}} C_{ba}^F(\delta f_a),\label{eqadjf}
\end{align}
\end{subequations}
where Eqs. (\ref{cnsvn}) and (\ref{cnsvpe}) represents the conservation property of the particle number, momentum, and energy, 
whereas Eqs. (\ref{eqadjt}) and (\ref{eqadjf}) are the self-adjointness of the operator. 
The other important property of Landau operator is Boltzmann's H-theorem,
\begin{eqnarray}
T_a\int d^3v \frac{\delta f_a}{f_{aM}}\left[C_{ab}^T(\delf_a)+C_{ab}^F(\delf_b) \right]\nonumber\\
+T_b\int d^3v \frac{\delta f_b}{f_{bM}}\left[C_{ba}^T(\delf_b)+C_{ba}^F(\delf_a) \right]\leq 0,\label{eqbolth}
\end{eqnarray}
where the equality is satisfied if and only if both $\delf_a$ and $\delf_b$ are perturbed Maxwellian,
\begin{equation}
\delf_a=f_{Ma}\left[\frac{\delta n_a}{n_a}+2\frac{\delta u_{\parallel a}\vpr}{v_a^2}+
\frac{\delta T_a}{T_a}\left(\frac{m_av^2}{2T_a}-\frac{3}{2}\right)\right],\label{eqfsm}
\end{equation}
while $\delta u_{\parallel b}=\delta u_{\parallel a}$ and $\delta T_b/T_b=\delta T_a/T_a$ for $\delf_b$.
Here, $v_a=\sqrt{(2T_a)/m_a}$ is the thermal velocity. 
Eq. (\ref{eqfsm}) corresponds to the lowest-order Taylor expansion of  shifted-Maxwellian with mean flow $\delta u_\parallel$
and density and temperature perturbations $\delta n$ and $\delta T$, 
\begin{equation}
f_{SMa}=(n_a+\delta n_a)\left(\frac{m_a}{2\pi (T_a+\delta T_a)}\right)^{3/2}\exp\left(-\frac{m_a(\vv-\delta\mathbf{u}_\parallel)^2}
{2(T_a+\delta T_a)}\right). \label{eqfsm2}
\end{equation}
It should be noted that the linearized collision operator for like-species collisions $C^T_{aa}+C^F_{aa}$ has another important 
property that the function of the form Eq. (\ref{eqfsm}) is the kernel of the like-species collision operator.
Since the drift-kinetic equation treats the distribution function which is already averaged over gyro-phase,
the perpendicular velocity moments such as $\int d^3v \vv_\perp \delf$ and $\int d^3v \vv_\perp C^L(\delf)$ vanish. 
Therefore, only the parallel component of the mean flow, $\delta u_\parallel$, appears in Eq. (\ref{eqfsm}).
In the same reason, only the parallel component of momentum balance relation in Eq. (\ref{cnsvpe}) is considered in the following sections. Note that $\vv_\perp$ represents the fast gyration motion around magnetic field and is 
different from the guiding-center drift velocity such as $E\times B$, $\nabla B$-, and curvature drift velocities,
in the drift-kinetic equation. The drift motion across the magnetic field lines is treated in the $\dot{\xx}(\dldl{}{\xx})$ term of Eq. (\ref{dke}).

It is known that the original linearized Landau operator does not satisfy the 
properties Eqs. (\ref{eqadjt}), (\ref{eqadjf}), and (\ref{eqbolth}) if $T_a\neq T_b$\cite{hinton}. Sugama's 
modified operator, Eq. (31) for $C^T_{ab}$ and Eq. (35) for $C^F_{ab}$ in Ref.\cite{sgm-c}, is made by modifying 
Landau operator so that it preserves the self-adjointness and H-theorem even if $T_a\neq T_b$. 
It is to be noted that if $T_a\neq T_b$, collisions between two background Maxwellian part 
$C_{ab}(f_{Ma},f_{Mb})$ is nonzero, which represents the thermalization 
process between two particle species. 
However, the main purpose of neoclassical transport simulation by the $\delf$-method is 
to evaluate the transport in a plasma with given kinetic profiles $n_a, n_b, T_a, T_b,\cdots$ and
we do not usually treat the change in background temperatures, which occurs very slowly in the transport
time scale compared to the collisional relaxation of perturbed distribution function to a quasi-steady state. 
Furthermore, the thermalization 
process between a light (a) and heavy (b) species occurs in a 
time scale $(m_a/m_b)(1-T_a/T_b)\tau_{ab}^{-1}$, which is much slower than the momentum relaxation 
time scale, $\tau_{ab}^{-1}=n_be_a^2e_b^2 \ln(\Lambda_{ab})(1+m_a/m_b)/(3\pi^{3/2}\epsilon_0^2m^2_a
(v_a^2+v_b^2)^{3/2})$ \cite{hinton}. It means that the large difference $T_a\neq T_b$
is allowed to happen only if the mass ratio of two species are very large,  such as between hydrogen and heavy
impurity ions, or between ions and electrons. Therefore, we ignore here the slow thermalization 
process on the background Maxwellian in neoclassical transport simulation. 
Our purpose is to construct a framework of drift-kinetic equation for $\delf$ using a linearized collision operator, 
which ensures the Onsager symmetry and H-theorem even if $T_a\neq T_b$. 
Momentum and energy exchange between two particle species occurs only through the $\delf$ part 
of the distribution function in this model.

\section{Implementation of Sugama operator in $\delf$ Monte Carlo code}\label{sec3}
Sugama's modified operator has already been implemented and benchmarked in continuum $\delf$ and full-$f$ gyrokinetic 
codes\cite{nnm,nkt,gt5dmi}, where the distribution function is discretized on the velocity space grids $(\vpr,\vpp)$, 
and the test- and field-particle operators are implemented numerically by finite-difference and numerical
integral schemes. In contrast, the $\delf$ Monte Carlo method is a particle code, in which the test-particle
operator is represented by the random walk of simulation markers in the velocity space, and the field-particle operator
is represented as a source/sink term on the markers' weight.
For the like-species collisions, FORTEC-3D has already implemented such a Monte Carlo scheme 
of operators $C^T_{aa}$ and $C^F_{aa}$. 
In this section, we show how the Sugama's operator is implemented for unlike-species collisions, $a\neq b$.

The modified test-particle operator is defined as follows:
\begin{align}
C^T_{ab}(\delf_a)&=C^{T0}_{ab}(\delf_a)+(\theta_{ab}-1)(\mathcal{P}_aC^{T0}_{ab}\delf_a
+C^{T0}_{ab}\mathcal{P}_a\delf_a)\nonumber\\
&+(\theta_{ab}-1)^2\mathcal{P}_aC^{T0}_{ab}\mathcal{P}_a\delf_a,\label{def_ctab}\\
  \theta_{ab}&\equiv\sqrt{T_a\left(\frac{1}{m_a}+\frac{1}{m_b}\right)/
\left(\frac{T_a}{m_a}+\frac{T_b}{m_b}\right)},\label{def_thab}
\end{align}
where $C^{T0}_{ab}$ represents the pitch-angle and velocity scattering terms, and it can be implemented by the 
random walk in the $(v=|\vv|, \xi=\vpr/v)$ space in Monte Carlo scheme\cite{boz,zlin}. 
The explicit form of $C^{T0}_{ab}$ in the Monte Carlo simulation is explained in \ref{apdx_a}. 
The projection operator $\mathcal{P}_a=\mathcal{P}_{1a}+\mathcal{P}_{2a}$ is defined as 
\begin{subequations}
\begin{align}
\mathcal{P}_{1a}f&\equiv f_{Ma}\frac{m_a}{T_a}\delta u_{\parallel a}[f]\vpr,\\
\mathcal{P}_{2a}f&\equiv f_{Ma}\frac{\delta T_a[f]}{T_a}\left(x^2_a-\frac{3}{2}\right),
\end{align}
\end{subequations}
where $x_a=v/v_a$, and 
\begin{subequations}
\begin{align}
\delta u_{\parallel a}[f]&\equiv \frac{1}{n_a}\int d^3v \vpr f, \label{defdu}\\ 
\frac{\delta T_a[f]}{T_a}&\equiv \frac{1}{n_a}\int d^3v \left(\frac{m_av^2}{3T_a}-1\right) f.\label{defdt}
\end{align}
\end{subequations}
Let us consider the change in the distribution function $\delf_a$ by a time
integral of short period $\Delta t$ according to $C^{T0}_{ab}$, i.e., $\delf_a^{(T0)}=
\Delta t C^{T0}_{ab}\delf_a^{(0)}$, where superscripts $(0)$ and $(T0)$ denote the distribution 
function before and after operating the random walk, respectively. 
By using the self-adjoint and particle conservation properties of the operator $C^{T0}_{ab}$, time integral of 
the terms on the RHS of Eq. (\ref{def_ctab}) can be approximately evaluated  as follows:
\begin{align}
\Delta t\mathcal{P}_aC^{T0}_{ab}\delf_a&=f_{Ma}\left[\frac{m_a\vpr}{T_a}\Delta u^{(T0)}_{ab}
+\frac{2}{3v_a^2}\left(x_a^2-\frac{3}{2}\right)\Delta E^{(T0)}_{ab}\right],\label{dtpct0}\\
\Delta t C^{T0}_{ab}\mathcal{P}_a\delf_a&=\Delta t\left[\frac{m_a}{T_a}C^{T0}_{ab}(\vpr f_{Ma})
\left\{\frac{\delta u_a^{(T0)}+\delta u_a^{(0)}}{2}\right\}\right.\nonumber\\
&\left.+C^{T0}_{ab}(x^2_af_{Ma})\left\{\frac{2}{3v_a^2}\left(
\frac{\delta E_a^{(T0)}+\delta E_a^{(0)}}{2}\right)-\delta n_a^{(0)}\right\}\right],\label{dtct0p}\\
\Delta t\mathcal{P}_aC^{T0}_{ab}\mathcal{P}_a\delf_a&=-\Delta t
\frac{4\hat{\nu}_{ab}\alpha_{ab}}{3\sqrt{\pi(1+\alpha_{ab}^2)}}f_{Ma}\left[\frac{m_a\vpr}{T_a}\left(
\frac{\delta u_a^{(T0)}+\delta u_a^{(0)}}{2}\right)\right.\nonumber\\
&\left. +\frac{2}{1+\alpha^2_{ab}}\left(x^2_a-\frac{3}{2}\right)\left\{\frac{2}{3v_a^2}\left(
\frac{\delta E_a^{(T0)}+\delta E_a^{(0)}}{2}\right)-\delta n_a^{(0)}\right\}
\right],\label{dtpct0p}
\end{align}
where 
\begin{subequations}
\begin{align}
\alpha_{ab}&=\frac{v_a}{v_b},\label{alpab}\\
\hat{\nu}_{ab}&=\frac{n_be_a^2e_b^2\ln\Lambda_{ab}}{4\pi \epsilon_0^2m^2_av_a^3},\label{nuab}\\
\delta n_a^{(0)}&=\frac{1}{n_a}\int d^3v~ \delf_a^{(0)}\left(=\frac{1}{n_a}
\int d^3v \delf_a^{(T0)}\right),\label{dnt0}\\
\delta u_{a}^{(0)}\left(\delta u_{a}^{(T0)}\right)&=\frac{1}{n_a}\int d^3v~ \vpr\delf_a^{(0)}
\left(\delf_a^{(T0)}\right),\label{dut0}\\
\delta E_{a}^{(0)}\left(\delta E_{a}^{(T0)}\right)&=\frac{1}{n_a}\int d^3v~ v^2\delf_a^{(0)}
\left(\delf_a^{(T0)}\right),\label{det0}\\
\Delta u^{(T0)}_{ab}&=\delta u_{a}^{(T0)}-\delta u_{a}^{(0)},\label{duabt0}\\
\Delta E^{(T0)}_{ab}&=\delta E_a^{(T0)}-\delta E_a^{(0)}.\label{deabt0}
\end{align}
\end{subequations}
Note here that $\delf_a^{(0)}$ and $\delta \{n, u, E\}_a^{(0)}$ mean the distribution function and the 
velocity moments before applying the $C^{T0}_{ab}$ operator, while $\delf_a^{(T0)}$ and 
$\delta \{u, E\}_a^{(T0)}$ mean those after applying $C^{T0}_{ab}$, respectively.
$\Delta u^{(T0)}_{ab}$ and $\Delta E^{(T0)}_{ab}$ represents the change in $\delta u_a$ and $\delta E_a$ by $C^{T0}_{ab}$. 
$\delta n_a^{(0)}$ is conserved in the $C^{T0}_{ab}$ operator.
Also note that the definition of $\hat{\nu}_{ab}$ in
the present paper equals to $3\pi^{1/2}/(4\tau_{ab})$ in Ref.\cite{sgm-c} 
where $\tau_{ab}=3\pi^{3/2}\epsilon_0^2m^2_av_a^3/(n_be_a^2e_b^2\ln\Lambda_{ab})$ in MKS unit.
The Coulomb logarithm $\ln\Lambda_{ab}$ can be generally defined for multi-species 
plasma as the ratio of the Debye length to the $90^\circ$ deflection impact parameter in the a-b collisions. 
In the present benchmark we will use a constant ($\ln\Lambda_{ab}=18$) for simplicity.
The symbol $\parallel$ in the subscript of $\delta u_a$ and $\Delta u_{ab}$ are omitted hereafter.
Since  the velocity moments such as Eqs. (\ref{dnt0})-(\ref{det0}) are easy to be evaluated in a 
particle $\delf$ code, numerical calculation of $C^T_{ab}$ 
in the form Eqs. (\ref{dtpct0})-(\ref{dtpct0p}) is a convenient way to implement the test-particle operator than to 
implement the original form of $C^T_{ab}$ by Sugama, Eqs (32) and (33) in Ref.\cite{sgm-c}, to a particle code.
Derivations of Eqs. (\ref{dtpct0})-(\ref{dtpct0p})  are explained in \ref{apdx_b}.

In the two-weight $\delf$ method for drift-kinetic simulations\cite{wang,brun}, 
distribution function is represented by the marker distribution 
function $g$ and the marker weights $(w, p)$ which satisfy the following relations:
\begin{eqnarray*}
\delf=wg=\sum_i w_i\delta(\mathbf{v}-\mathbf{v}_i)&,&f_M=pg=\sum_i p_i\delta(\mathbf{v}-\mathbf{v}_i),
\end{eqnarray*}
where the subscript $i$ represents the index of simulation markers and $\delta(\mathbf{v})$ is the Dirac $\delta$ function,
respectively. Consider here the Monte Carlo operator 
$\Delta t C^{T0}_{ab}$  changes each marker's velocity $\mathbf{v}_i\rightarrow \mathbf{v}_i+\Delta
\mathbf{v}_i$. Then, the change of $\delf_a$ by the whole test-particle part can be expressed formally as follows: 
\begin{eqnarray}
\delf_a^{(T0)}(\mathbf{v}_i+\Delta\mathbf{v}_i)&=&\delf_a^{(0)}(\mathbf{v}_i)+
f_{Ma}(\mathbf{v}_i+\Delta\mathbf{v}_i))\times\nonumber\\
&&S^T_{ab}\left[\delta n_a^{(0)},\delta u_a^{(0)},\delta E_a^{(0)},\delta u_a^{(T0)},\delta E_a^{(T0)},\theta_{ab};\vv_i+\Delta\vv_i\right],\label{dfdtct}
\end{eqnarray}
where $S^T_{ab}$, which is a functional of velocity moments of $\delf_a$ before and after applying the
random walk $C^{T0}_{ab}$, represents the three terms which is proportional to $(\theta_{ab}-1)$ 
in Eq.(\ref{def_ctab}), and we have used the fact that not only Eqs. (\ref{dtpct0}) and (\ref{dtpct0p}) but also
$C^{T0}_{ab}(\vpr f_{Ma})$ and $C^{T0}_{ab}(x^2_af_{Ma})$ 
in Eq. (\ref{dtct0p}) are analytic functions which are proportional to $f_{Ma}(\vv)$ (See \ref{apdx_b}). 
According to the source term $S^T_{ab}$, each marker weight $w_i$ changes by $\Delta t C^T_{ab}$ as 
\begin{equation}
w_i^{(T)}=w_i^{(0)}+(f_{Ma}/g)S^T_{ab}=w_i^{(0)}+p_iS^T_{ab}.\label{dwdtct}
\end{equation}
Note that the weight $p_i$ does not change by the linearized collision operator, and $S^T_{ab}$ vanishes
if $T_a=T_b~ (\theta_{ab}=1)$.  
Hereafter, we represent the distribution function after 
operating $C^T_{ab}$ as $\delf_a^{(T)}$.\\

Next, let us consider the field-particle operator $C^F_{ab}$.  Sugama's modified Landau operator 
for the field-particle term is made so as to satisfy both the conservation and self-adjoint properties, as follows:
\begin{align}
\Delta t C^F_{ab}(\delf_b)&=f_{Ma}(v)\left[c_0\left(\frac{1}{n_a}-\frac{3Q_{ab}}{2}\right) \right. \nonumber\\
&\left. \qquad+c_1\delta V_{ba}^T R_{ab}(v,\vpr)+c_2\delta W_{ba}^T Q_{ab}(v)\right],
\label{eqcf}\\
\delta V_{ba}^T&=\frac{1}{m_av_a}\int_t^{t+\Delta t}dt \int d^3v\frac{\delf _b}{f_{Mb}}
 C^T_{ba}(m_b\vpr f_{Mb})\nonumber\\
&=\frac{1}{m_av_a}\int_t^{t+\Delta t}dt \int d^3v m_b\vpr C^T_{ba}(\delf_b)=
\frac{n_bm_b}{m_av_a}\Delta u^{(T)}_{ba},\label{eqdv}\\
\delta W_{ba}^T&=\frac{1}{T_a}\int_t^{t+\Delta t}dt \int d^3v \frac{\delf_b}{f_{Mb}} C^T_{ba}
\left(\frac{m_bv^2}{2}f_{Mb}\right)\nonumber\\
&=\frac{1}{T_a}\int_t^{t+\Delta t}dt \int d^3v \frac{m_bv^2}{2} C^T_{ba}(\delf_b)=
\frac{n_bm_b}{2T_a}\Delta E^{(T)}_{ba},\label{eqdw}\\
R_{ab}(v,\vpr)&=f_{Ma}(v)^{-1}\frac{C^T_{ab}\left(\frac{\vpr}{v_a}f_{Ma}\right)}
{\int d^3v \frac{\vpr}{v_a}C^T_{ab}\left(\frac{\vpr}{v_a}f_{Ma}\right)}\nonumber\\
&=\frac{2\theta_{ab}(1+\alpha_{ab}^2)^{5/2}}
{n_a\alpha_{ab}^3\left(m_a/m_b+1\right)}\left(\frac{\vpr}{v_a}\right)
\left[\frac{3\sqrt{\pi}G(x_b)}{2x_a}+\frac{\alpha_{ab}(\theta_{ab}-1)}{(1+\alpha_{ab}^2)^{3/2}}\right],\\
Q_{ab}(v)&=f_{Ma}(v)^{-1}\frac{C^T_{ab}(x_a^2f_{Ma})}{\int d^3v x^2_a C^T_{ab}(x_a^2f_{Ma})}\nonumber\\
&=\frac{2\theta_{ab}(1+\alpha_{ab}^2)^{5/2}}
{3n_a\alpha_{ab}^3\left(m_a/m_b+1\right)}\left[\frac{3\sqrt{\pi}}{4\alpha^2_{ab}x_a}\left\{\Phi(x_b)-x_b\Phi'(x_b)
(1+\alpha_{ab}^2)\right\}\right.\nonumber\\
&\qquad+\left. \frac{\alpha_{ab}(\theta_{ab}-1)}{(1+\alpha_{ab}^2)^{3/2}}\left(x_a^2-\frac{3}{2}\right)\right] ,
\end{align}
where 
\begin{subequations}
\begin{align}
\Delta u^{(T)}_{ba}&=\delta u_{b}^{(T)}-\delta u_{b}^{(0)},\\
\Delta E^{(T)}_{ba}&=\delta E_b^{(T)}-\delta E_b^{(0)},
\end{align}
\end{subequations}
represent the change in parallel mean flow and energy of particle species $b$ by the test-particle collisions 
$C^{T}_{ba}$ respectively, and $\Phi(x)$ and $G(x)$ are defined in \ref{apdx_a}.
We have modified the form of Eq. (\ref{eqcf}) from the original one, Eq. (35) in Ref.\cite{sgm-c}. The term proportional to $c_0$
is introduced to ensure the particle-number conservation property in $C^T_{ab}+C^F_{ab}$. 
Also, the self-adjointness of $C^T$ is used to derive Eqs. (\ref{eqdv}) and (\ref{eqdw}).
In the ideal limit where there is no numerical error in evaluating 
velocity and time integrals in $C^T_{ab}$, 
$(c_0,c_1,c_2)=(0,-1,-1)$ corresponds to the original form of $C^F_{ab}$ in Ref.\cite{sgm-c}. 
However, as it has been pointed out in previous studies on like- and 
unlike-species collision operators\cite{f3d1,nkt,wang,koles}, direct numerical implementation of $C^F_{ab}$ as in the 
original form fails to keep the conservation properties of collision operator because of the numerical errors 
in the velocity and time integrals. 
Therefore, as we have adopted for like-species collision operator in FORTEC-3D\cite{f3d1} 
and in a continuum gyrokinetic full-f code GT5D\cite{idm}, 
the numerical factors $(c_0,c_1,c_2)$ are determined at each time when collision term is operated
so that it ensures the conservation of particle number, momentum, and energy in the collisions 
between particle species $a$ and $b$, i.e., Eqs. (\ref{cnsvn}) and (\ref{cnsvpe}). This procedure is done 
as follows. First, consider the time integral of the conservation law Eqs. (\ref{cnsvn}) and (\ref{cnsvpe}) over short 
time step $\Delta t$, and substitute the expression of $C^{F}_{ba}$ Eq. (\ref{eqcf}) to these equations. It becomes
\begin{eqnarray}
\Delta t \int d^3v 
\begin{pmatrix}
C^{T}_{ba}(\delf_b) \\ m_a \vpr C^{T}_{ab}(\delf_a)\\ m_a v^2C^{T}_{ab}(\delf_a)
\end{pmatrix}
=\begin{pmatrix}
\delta N^{T}_{ba} \\ m_b v_b \delta V^T_{ab} \\ 2T_b\delta W^T_{ab}
\end{pmatrix}• 
=-\Delta t \int d^3v C^{F}_{ba}(\delf_a)
\begin{pmatrix}
1 \\ m_b \vpr \\ m_b v^2
\end{pmatrix}\nonumber\\
=-\int d^3v f_{Mb}
\begin{pmatrix}
\left(\frac{1}{n_b}-\frac{3Q_{ba}}{2}\right) & R_{ba} & Q_{ba} \\
m_b\vpr \left(\frac{1}{n_b}-\frac{3Q_{ba}}{2}\right) & m_b\vpr R_{ba} & m_b\vpr Q_{ba} \\
m_b v^2\left(\frac{1}{n_b}-\frac{3Q_{ba}}{2}\right) & m_b v^2R_{ba} & m_b v^2Q_{ba} 
\end{pmatrix}
\cdot\begin{pmatrix}
c_0 \\
c_1 \delta V^T_{ab} \\
c_2 \delta W^T_{ab} 
\end{pmatrix} ,\label{eqcx}
\end{eqnarray}
where
\begin{equation*}
\delta N_{ba}^T=\int_t^{t+\Delta t}dt \int d^3v C^T_{ba}(\delf_b)= \int d^3v[\delf_b^{(T)}-\delf_b^{(0)}].
\end{equation*}
Note here that it is not $\delta N^T_{ab}$ but $\delta N^T_{ba}$ which should appear in Eq.(\ref{eqcx}) so that 
the error in particle number conservation in $C^T_{ba}$ is compensated by $C^F_{ba}$.
The velocity integral of the term $f_{Mb}R_{ba}$, $f_{Mb}Q_{ba}$ etc. are numerically carried out
by utilizing the relation of marker weight $p_ig=f_{M}(v_i)$ as follows,
\begin{eqnarray}
\int d^3v f_{Mb}(v)R_{ba}(v,\vpr)\simeq \sum_i p_i R_{ba}(v_i,v_{\parallel,i}),
\end{eqnarray}
and so on. Then, by inverting the $3\times 3$ matrix of which components are evaluated numerically as above, 
we have a set of simultaneous equations to determine the coefficients $(c_0,c_1,c_2)$ which satisfies the 
conservation laws. The validity of the modified $C^F_{ab}$ operator
for unlike-species collisions will be confirmed later in the benchmark calculations. 

Since Eq. (\ref{eqcf}) is proportional to $f_{Ma}$, it can be rewritten as $S^F_{ab}f_{Ma}$. Therefore, the field-particle
operator act as another source term on the marker weight $w_i$ as in a similar way as $S^T_{ab}$ from the test-particle part.
Therefore, in summary, after operating the $C^T_{ab}$ and $C^F_{ab}$ terms, each marker's 
velocity $\mathbf{v}_i$ and weight $w_i$ change as follows:
\begin{eqnarray*}
\mathbf{v}_i(t+\Delta t)&=&\mathbf{v}_i(t)+\Delta\mathbf{v}_i,\\
w_i(t+\Delta t)&=&w_i^{(T)}+p_iS^F_{ab}=w_i^{(0)}+p_i(S^T_{ab}+S^F_{ab}).
\end{eqnarray*}
Note here that the collision operator should be evaluated and operated in the order 
$C^{T0}\rightarrow S^T \rightarrow S^F$
in the $\delf$ Monte Carlo simulation. Since $S^F_{ab}$ requires information of change in momentum and
energy of the opponent particle species $b$, $C^L_{ab}(\delf_a)$ and $C^L_{ba}(\delf_b)$ should be 
calculated simultaneously. If there are three or more particle species, one needs to repeat the procedure
for all combinations including like-species collisions, (a,a), (a,b), (a,c), $\cdots$, (b,b), (b,c), $\cdots$ etc.

\section{Benchmark of the collision operator}\label{sec4}
\subsection{Benchmark with a full-f PIC code}\label{sec41}
To benchmark the modified linearized collision operator, firstly we compared the momentum and 
energy relaxation process between two ion species simulated by a different model of collision operator.
Here we used a full-f PIC simulation code\cite{pianpic} with Nanbu-model collision operator\cite{nmb} to benchmark
our $\delf$ Monte Carlo code.
In a full-f PIC code, the total distribution function $f_a=f_{Ma}+\delf_a$ rather than the perturbed part $\delf_a$ 
is represented by the simulation markers, of which weights are usually uniform. 
This is the main difference from the $\delf$ code in which markers have non-uniform weight. 
To simulate the Coulomb collision process by Nanbu model operator, it  randomly chooses pairs of 
simulation markers in a tiny cell, and then evaluate the cumulative scattering angle of these two 
charged particles from the impact parameter and relative velocity.
This collision operator is essentially nonlinear, while the linearized collision term neglects the 
$C_{ab}(\delf_a,\delf_b)$ term.

The benchmark has been carried out as follows. The initial distribution functions of particle species 
$a$ and $b$ in the $\delf$ code were given as shifted Maxwellian as Eq. (\ref{eqfsm}) with different initial 
perturbation amplitude $\delta u_a/v_a=0.1,\ \delta u_b/v_b=-0.1,\
\delta T_a/T_a=0.05,\ \delta T_b/T_b=-0.05$. 
Note that in the $\delf$ code, the initial simulation marker distribution $g$ is given by random sampling of $f_M$ and therefore
the marker weight of particle species $a$ is uniform, $p_a=f_{Ma}/g=n_a/N$, where $N$ is the number of simulation markers. 
The initial shifted Maxwellian is then represented by non-uniform initial weight $w_{ai}$ : 
\begin{eqnarray*}
w_{ai}(t=0)=p_a\left[\frac{\delta n_a}{n_a}+2\frac{\delta u_av_{\parallel i}}{v_{a}^2}+
\frac{\delta T_a}{T_a}\left(\frac{m_av^2_i}{2T_a}-\frac{3}{2}\right)\right],
\end{eqnarray*}
On the other hand, in the full-f PIC code, the initial simulation particles are loaded by the random sampling of 
the shifted Maxwellian in the form of Eq. (\ref{eqfsm2}).
The difference in the initial distribution between the $\delf$ code and the full-f PIC code is negligibly small, 
$\mathcal{O}(\delta^2)$ where $\delta\sim (\delta n_a/n_a), (\delta u_a/v_a), (\delta T_a/T_a)$.

To test the mass-ratio dependence, the particle 
species $a$ was chosen as H$^+$ while $b$ was varied (D, T, C, Ar, Fe, Kr, Ag, W). The ion charge of 
species $b$ and $\ln\Lambda_{ab}$ were fixed to $+2$ and 18 respectively, for simplicity.
The other parameters were : $n_a=1\times 10^{19}$[m$^{-3}$],  $n_b=0.5\times 10^{19}$[m$^{-3}$],
$T_a=T_b=2$[keV], $\delta n_a/n_a=\delta n_b/n_b=0.1$. Note here that the Nanbu operator in
full-f PIC code contains the slow energy equilibration process among two Maxwellians $C_{ab}(f_{Ma},f_{Mb})$
if $T_a\neq T_b$, while it is neglected in the $\delf$ code. Therefore, we compared here the two simulation
codes in $T_a=T_b$ case.  
In these benchmark calculations, $N_m=2\times 10^5$ markers per one species were used in both two codes. 
The time step size in the $\delf$ code is determined by $\tau_{min}$, which is the minimum 
value of $\hat{\nu}_{ij}^{-1}(i, j=a$ or $b)$ in each case. 
In the present benchmarks, $\Delta t=5\times 10^{-4}\tau_{min}$ was chosen. 
In the following, the correction scheme  in the field-particle operator was used unless otherwise noted.  

\begin{figure}[tb]
  \includegraphics[width=7cm]{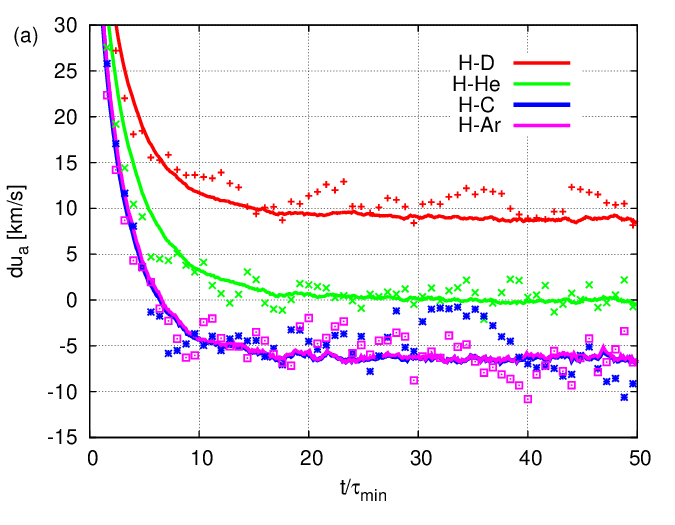}
  \includegraphics[width=7cm]{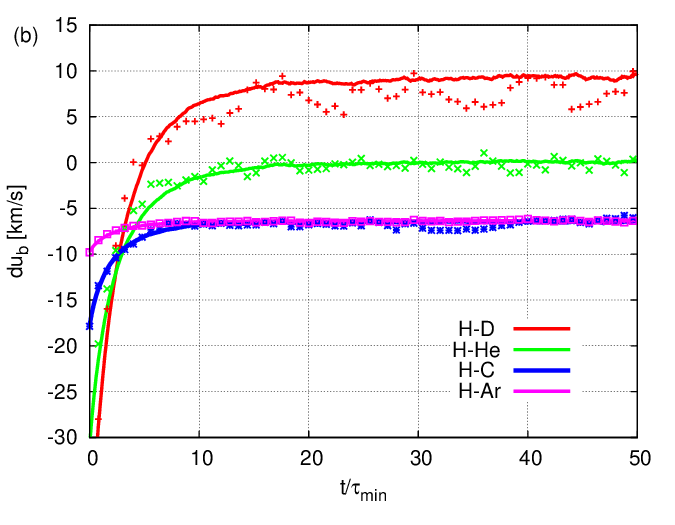}
  \includegraphics[width=7cm]{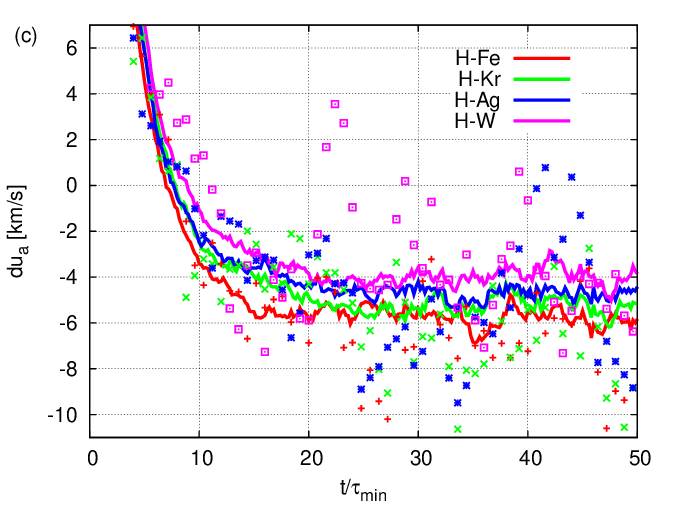}
  \includegraphics[width=7cm]{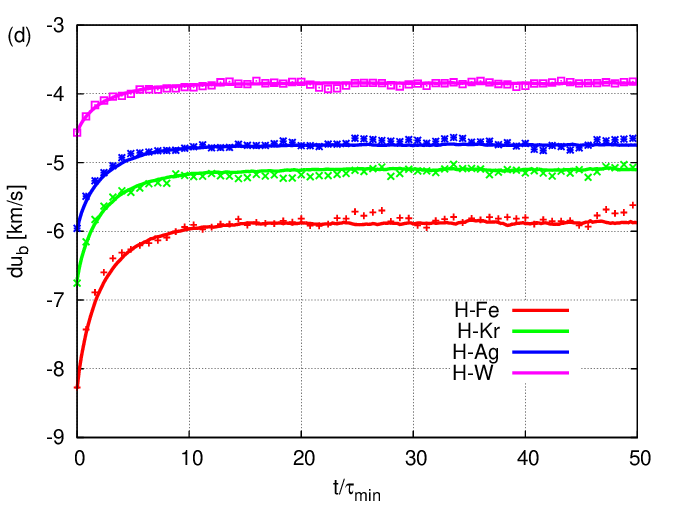}

\vspace{1cm}
  \caption{Time evolution of parallel mean flow $\delta u$ in two ion species  plasmas, $a=$H$^+$ and $b=$
(D, T, $\cdots$, W)$^{2+}$. Figures (a) and (b) are the results of lighter $b$ species,
while (c) and (d) are heavier $b$ species, respectively. Simulation results from the $\delf$
code with linearized collision operator are shown by solid curves and those from full-f PIC with 
Nanbu operator are plotted by points. Horizontal axis is time normalized by collision time 
$\tau_{min}$, which is the minimum value of $\hat{\nu}_{ij}(i, j=a$ or $b)$ in each case.}
  \label{fig:1}
\end{figure}

\begin{figure}[tb]
  \includegraphics[width=7cm]{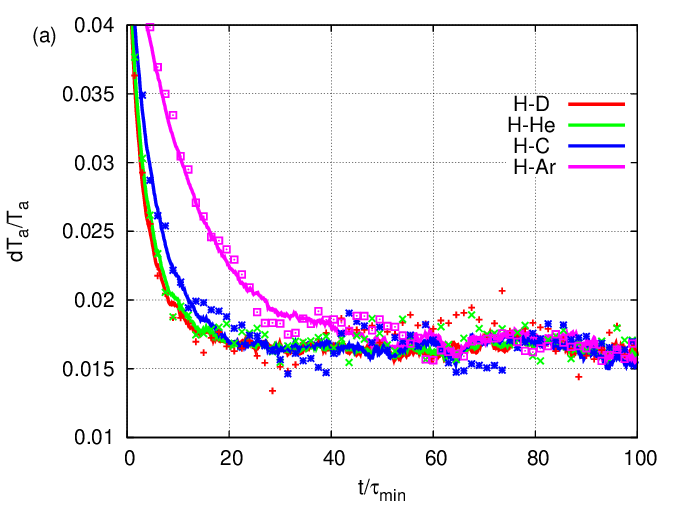}
  \includegraphics[width=7cm]{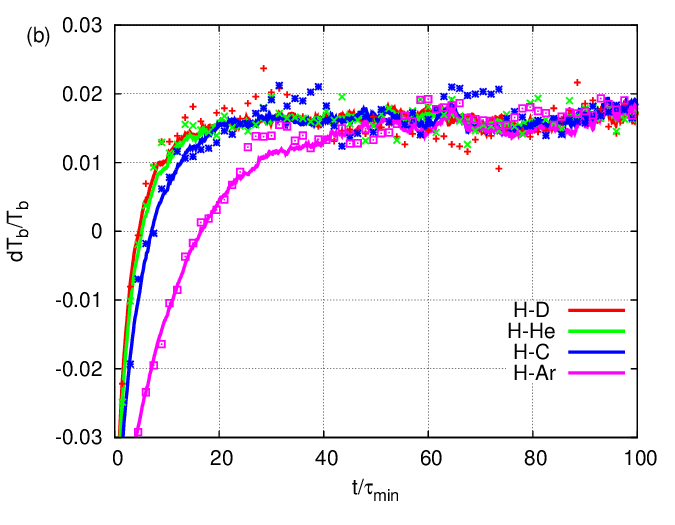}
  \includegraphics[width=7cm]{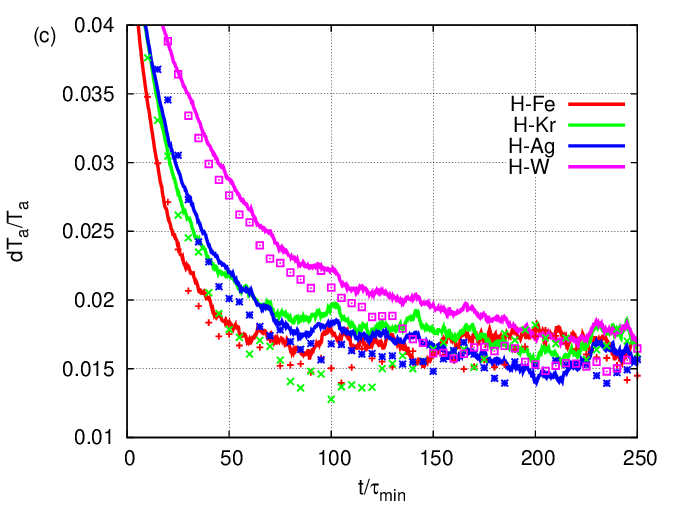}
  \includegraphics[width=7cm]{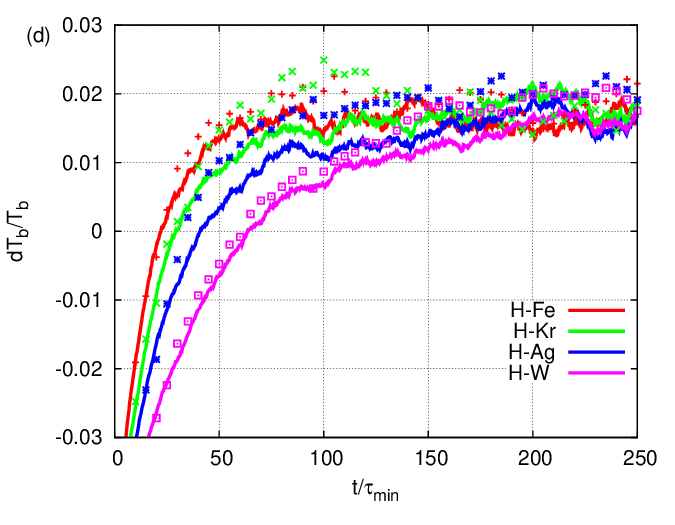}

\vspace{1cm}
  \caption{Time evolution of temperature  perturbation $\delta T/T$ in two ion species  plasmas.
Simulation results from the $\delf$ code with linearized collision operator are shown by solid curves 
and those from full-f PIC with Nanbu operator are plotted by points.}
  \label{fig:2}
\end{figure}

Figures \ref{fig:1} and \ref{fig:2} show the time evolution of the parallel mean flow $\delta u$ and 
temperature fluctuation $\delta T/T$ as defined in Eqs. (\ref{defdu}) and (\ref{defdt}). 
It was found that the damping rate 
of $\delta u$ and $\delta T/T$ of both species $a$ and $b$ agreed well between $\delf$ Monte Carlo code and
the full-f PIC code. One can see that the final stationary state satisfies the condition $\delta u_a\simeq \delta u_b$
and $\delta T_a/T_a\simeq \delta T_b/T_b$, as it is expected from the H-theorem. 
Here, the expectation values 
of  $\delta u$ and $\delta T/T$ at $t\rightarrow \infty$ can be evaluated from the initial values, by 
using the conservation of total momentum and energy, as follows:
\begin{eqnarray}
\delta u(t\rightarrow \infty)&=&\frac{\sum_a m_a n_a \delta u_a(t=0)}{\sum_a m_a n_a},\nonumber\\
\frac{\delta T}{T}(t\rightarrow \infty)&=&\frac{\sum_a n_a \delta T_a(t=0)}{\sum_a n_a T_a}.\label{exp_ut}
\end{eqnarray}
The convergence of the $\delf$ simulation was checked by varying the number of simulation markers as 
shown in Table \ref{tbl:1} for the H-Fe case.    
It was confirmed that the $\delta u$ and $\delta T/T$ in the simulations converged to their expectation values, and 
the standard deviation was roughly proportional to $1/\sqrt{N_m}$.
Therefore, we conclude that both simulation methods correctly simulate the damping process of distribution 
function towards the stationary state. 
The statistic noise on $\delta u$ and $\delta T/T$ is larger in the full-f simulation than that in
$\delf$ simlation. The fluctuation is caused by the sampling noise of distribution
function by finite number of markers. Since the markers in $\delf$ method is used for sampling only the 
small perturbation part of distribution function, $\delf$, it is expected that the sampling noise level in
a $\delf$ simulation is $\mathcal{O}[(\delf/f_{M})^2]$ smaller than that in a full-f method if they 
use the same number of simulation markers\cite{xukrom}.  This is the advantage of the $\delf$ scheme. 
However, the $\delf$ method cannot be applied to a far non-equilibrium system where $\delf/f_{M}\sim\mathcal{O}(1)$ 
such as SOL/divertor region of torus plasmas, and the full-f method is indispensable to simulate the kinetic
transport process in such a case. 

\begin{table}[htb]
  \begin{center}
    \caption{Time average and standard deviation of $\delta u$ and $\delta T/T$ (H-Fe case) 
evaluated from t=175$\tau_{min}$ to 250$\tau_{min}$. $N_m$ is the number of simulation markers.
The theoretical expectation values are obtained from Eq. (\ref{exp_ut}).}
    \begin{tabular}{|c||c|c|c|c|} \hline
$N_m$ & $\delta u_H$[km/s] & $\delta u_{Fe}[km/s]$ & $\delta T_H/T_H (\times 100)$ & $\delta T_{Fe}/T_{Fe} (\times 100)$ \\ \hline \hline
48000    & -5.545$\pm$ 1.717 & -5.832$\pm$ 0.061 & 1.834$\pm$ 0.099 & 1.399$\pm$ 0.197 \\ \hline
192000  & -5.903$\pm$ 0.821 & -5.874$\pm$ 0.029 & 1.711$\pm$ 0.057 & 1.583$\pm$ 0.114 \\ \hline
768000  & -5.962$\pm$ 0.381 & -5.850$\pm$ 0.014 & 1.649$\pm$ 0.031 & 1.679$\pm$ 0.062 \\ \hline
3072000 &-5.800$\pm$ 0.191 & -5.856$\pm$ 0.007 & 1.690$\pm$ 0.032 & 1.662$\pm$ 0.063 \\ \hline
Expectation &  -5.852 & -5.852 & 1.667 & 1.667 \\ \hline
     \end{tabular}\label{tbl:1}
  \end{center}
\end{table}

\begin{figure}[tb]
  \includegraphics[width=7cm]{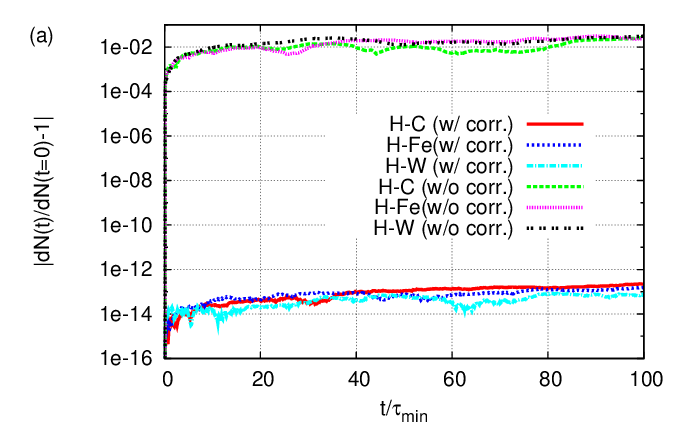}
  \includegraphics[width=7cm]{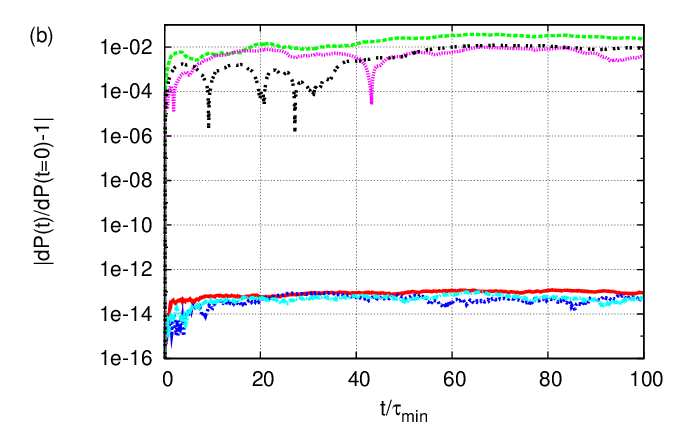}
  \includegraphics[width=7cm]{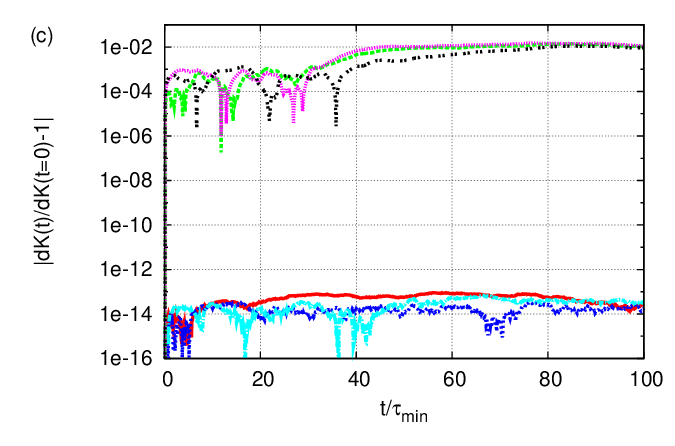}
\vspace{1cm}
  \caption{Relative error in the (a) total particle number $\delta N$, 
(b) parallel momentum $\delta P$, and (c) kinetic energy $\delta K$ from 
 their initial values in the case of $a=$H$^+$ and $b=($C, Fe, W$)^{2+}$ plasmas. 
``w/ corr.'' curves represent the calculations with the correction scheme 
 in $C^F_{ab}$ term to keep the conservation property, 
while ``w/o corr.'' curves represent the calculations without correction, i.e.,  $(c_0,c_1,c_2)=(0,-1,-1)$ in Eq. (\ref{eqcf}).}
  \label{fig:3}
\end{figure}

\begin{figure}[tb]
  \includegraphics[width=7cm]{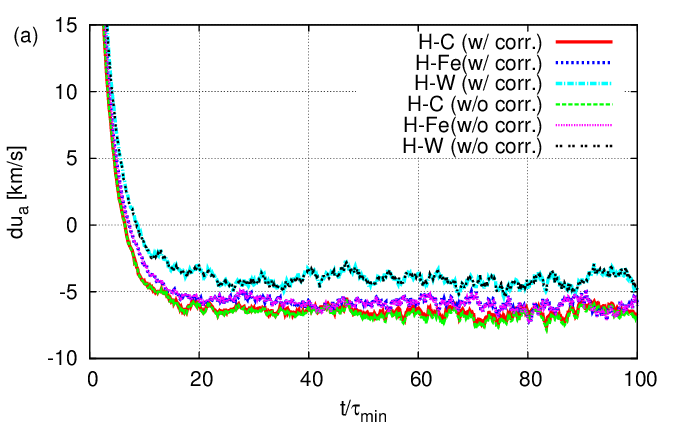}
  \includegraphics[width=7cm]{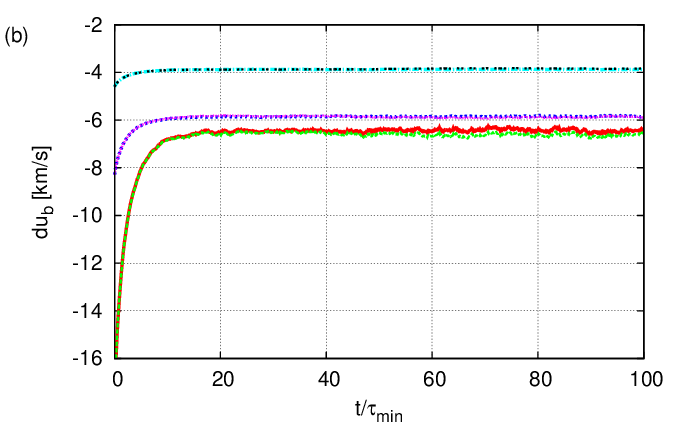}

\vspace{1cm}
  \caption{Time evolution of parallel mean flow, (a): $\delta u_a$ and (b): $\delta u_b$ in the case of 
$a=$H$^+$ and $b=($C, Fe, W$)^{2+}$ plasmas with or without the correction scheme in the $C^F_{ab}$ term 
to keep the conservation property. }
  \label{fig:4}
\end{figure}

In figures \ref{fig:1} and \ref{fig:2}, it is found that  the momentum transfer 
between two species completes  in the time scale of $10\tau_{min}$, while the energy equilibration time  
becomes slower as the mass ratio $m_b/m_a$ is larger. 
This tendency can be explained as follows. In \ref{apdx_a}, Langevin equations for the change of $v^2$ and $\xi=\vpr/v$
by $C^{T0}_{ab}$ are derived. $\gamma^{ab}_{v^2}(v)$ and $\gamma^{ab}_{\xi}(v)$ appeared in Eqs.(\ref{a5}) and (\ref{a6})
represent the damping rates of energy and parallel momentum by collisions, respectively. In the case $m_a\ll m_b$, 
$\alpha_{ab}=v_a/v_b\sim\sqrt{m_b/m_a}\gg 1$ and $\gamma^{ab}_{v^2}/\gamma^{ab}_{\xi}\simeq 2/\alpha_{ab}^2\ll 1$ 
for $v\gtrsim v_a$ particles.
Therefore, the energy equilibration becomes relatively slower than the momentum transfer as $\alpha_{ab}$ increases.
Note here that the ratio  $\gamma^{ab}_{v^2}/\gamma^{ab}_{\xi}$ is independent of the ion charge $e_a$ and $e_b$ but 
depends on $\alpha_{ab}$. As discussed in \ref{apdx_a}, the simulation time step size $\Delta t$ should be chosen so that 
$\alpha_{ab}\Delta t\hat{\nu}_{ab}\ll 1$ is satisfied, otherwise the energy diffusion term in
the test-particle operator has a numerical problem.
Therefore, when one carries out a neoclassical transport simulation which includes several ion species with separated masses,  
we need to set smaller time step size as $\alpha_{ab}$ increases while the energy equilibration becomes slower and slower. 
Then, it will take many time steps until it reaches a
quasi-steady state, and the  accumulation of numerical error in 
particle number $\delta N\equiv \sum_a|\delta n_a|$, total momentum 
$\delta P\equiv \sum_a m_an_a\delta u_a$, and total kinetic energy $\delta K\equiv\sum_a (m_an_a/2)\delta E_a$
in the long-term simulation might be a problem. Here, $\delta n$, $\delta u$, and $\delta E$ are defined in 
Eqs. (\ref{dnt0})-(\ref{det0}). Strictly speaking, the conservation of particle numbers should 
be satisfied in each particle species $\delta n_a$ independently. However,  we check the conservation 
property of particle number by summing up the absolute values of $\delta n_a$ for simplicity.  
Figure \ref{fig:3} shows the time evolution of accumulated relative error in $|\delta N|$, $|\delta P|$, and 
$|\delta K|$ in the $\delf$ simulation for the three cases $(a=H,b=$C or Fe or W$)$. If the 
correction scheme in the field-particle operator is turned on, the coefficients $(c_0,\ c_1,\ c_2)$ in Eq. (\ref{eqcf}) are 
determined so that the linearized collision operator $C^T_{ab}+C^F_{ab}$ should satisfy the 
conservation properties, Eqs. (\ref{cnsvn}) and (\ref{cnsvpe}), among the all combinations of  two particle species including 
like-species collisions. In Figure \ref{fig:3}, it is clearly demonstrated that the correction scheme kept the relative error in
$|\delta N|$, $|\delta P|$, and $|\delta K|$ within the rounding-error level, and no accumulation of 
numerical error happened up to 100 collision times.
On the other hand, if the correction scheme was turned off, the relative error in conserved quantities 
became $\mathcal{O}(0.01)$. Though the numerical error level can be reduced by increasing the 
number of simulation markers as shown in Table \ref{tbl:1}, the correction scheme suppresses the numerical 
error very efficiently.
Figure \ref{fig:4} is the comparison of the time evolution of $\delta u$ between with and without the correction term in 
$C^F_{ab}$. It demonstrates that the correction term has no side effect on the time evolution of mean
flow towards the steady state solution. We have also checked that the correction scheme does not
affect the time evolution of $\delta T/T$.  

\begin{figure}[tb]
  \includegraphics[width=7cm]{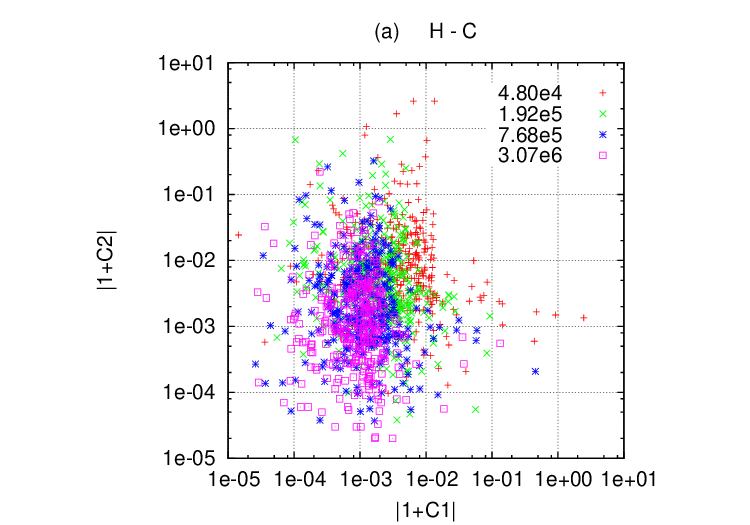}
  \includegraphics[width=7cm]{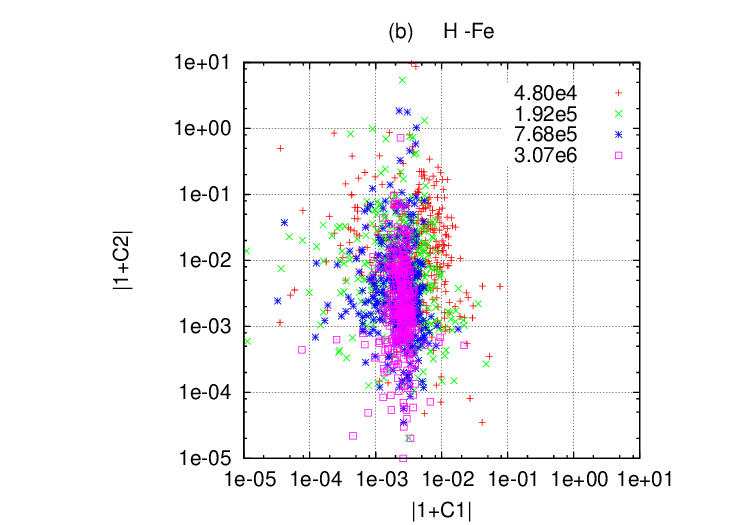}
  \includegraphics[width=7cm]{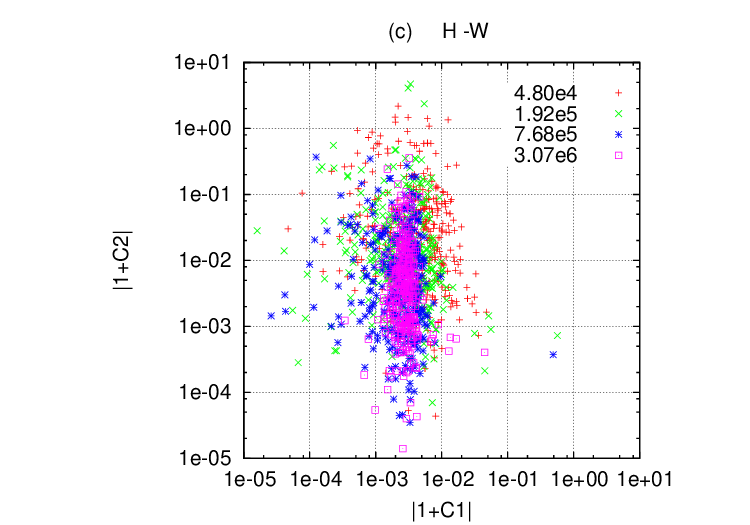}
\vspace{1cm}
  \caption{The distribution of the correction coefficients $c_1$ and $c_2$ in $C^F_{ab}$ term 
in the case of  $a=$H$^+$ and $b=($C, Fe, W$)^{2+}$ plasmas. Total number of simulation markers (written in the 
legend in the figures) are varied in these 4 cases. $|1+c_{1,2}|\rightarrow 0$ is the ideal limit 
if the numerical error vanishes completely in the velocity integrals.}
  \label{fig:5}
\end{figure}

It should be noted that the correction scheme compels the $C^F_{ab}$ 
operator to satisfy the conservation properties of linearized operator, even if there is any mistake 
or large error in $C^T_{ab}$ or $C^F_{ab}$ themselves. To verify the correction scheme,
we investigated how the correction coefficients $c_i$ in Eq. (\ref{eqcf}) behaved.
In Figure \ref{fig:5}, the distribution of $(|1+c_1|,|1+c_2|)$ for $C^F_{ab}$ during the simulations 
for $(a,b)=$(H,C), (H, Fe), and (H,W) plasmas are shown. In the ideal limit $c_1$ and $c_2$ should 
converge to -1. As the number of simulation markers increases,  $(|1+c_1|,|1+c_2|)$  was found to approach 
to $(0,0)$ as it is expected. 
We also found that the coefficient $c_0$ also remained small, $|c_0|<10^{-4}$, 
during the simulations.  Therefore, it is verified that the collision operators $C^T_{ab}$ and $C^F_{ab}$
are correctly implemented. 
As the mass ratio $m_b/m_a$ becomes larger, $|1+c_2|$ tends to spread in wider 
range $0.0001\sim 0.1$ than $|1+c_1|$ even if we use many simulation markers.  It is inferred that the 
statistic noise from the random-walk part in $C^{T0}_{ab}$ tends to affect the conservation property 
when $m_a/m_b\ll 1$. We will discuss on this point later in Summary.\\

\subsection{Benchmark with a continuum gyrokinetic $\delf$ code}\label{sec42}
Next, to check the self-adjoint property of the collision operator and the H-theorem in the case of 
different temperatures, we carried out simulations of a 4-species plasma
(D$^+$, T$^+$, He$^{+2}$, C$^{+6}$) by the $\delf$ Monte Carlo code and compared the result with the 
same Sugama operator implemented in a continuum gyrokinetic $\delf$ code\cite{nkt}.
The initial plasma parameters  are shown in Table \ref{tbl:2}. 
The density of each species is chosen so that they have similar amplitude of initial parallel momentum
$|m_a n_a\delta u_a|$. Time step size in the $\delf$ Monte Carlo simulation 
was set to $\Delta t=2.5\times 10^{-4} \tau_{min}$ where $\tau_{min}=1/\hat{\nu}_{CC}$ in this case, and the
number of simulation markers per species was $N_m=4.8\times 10^5$. 

\begin{figure}[tb]
  \includegraphics[width=7cm]{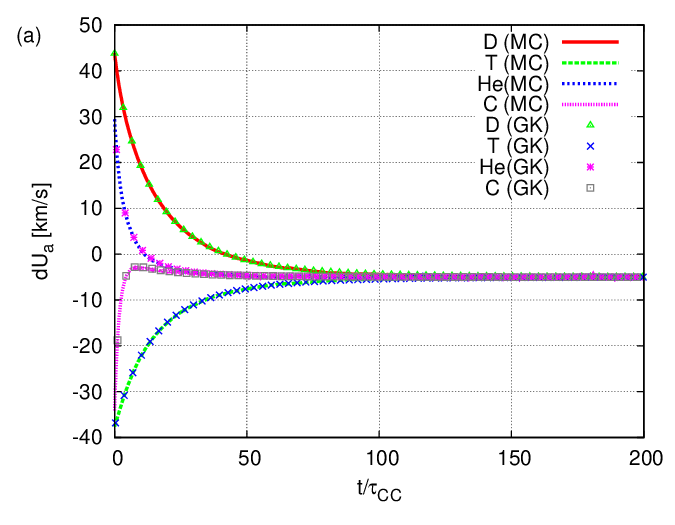}
  \includegraphics[width=7cm]{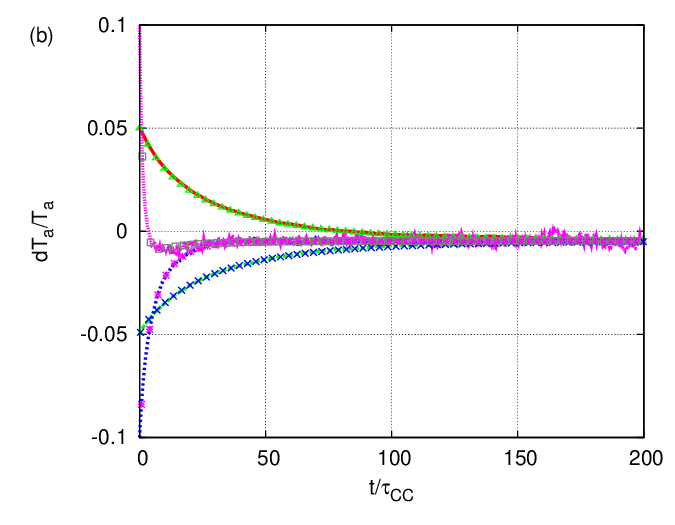}

\vspace{1cm}
  \caption{Time evolution of (a): mean flow $\delta u$ and (b) : temperature  perturbation $\delta T/T$ 
in the 4-species plasma, of which initial condition was given as in Table \ref{tbl:2}.
Simulation results from the $\delf$ Monte Carlo code (MC) are plotted by curves, 
and those from the continuum gyrokinetic $\delf$  code (GK) are plotted by points. Note that the MC simulations 
shown here did not apply the weight averaging method which is discussed in Figures \ref{fig:7}.}
  \label{fig:6}
\end{figure}

First, let us compare the time evolution of mean flow $\delta u$ and the temperature perturbation
$\delta T/T$ from the two codes. As shown in Figure \ref{fig:6}, the damping rate and the final steady state values of
$\delta u$ and $\delta T/T$ agreed very well between two codes. It took $t\sim150\hat{\nu}_{CC}^{-1}$ to reach to 
a stationary state where all the $\delta u_a$ and $\delta T_a/T_a$ of four species converge to the same value, as 
it is expected. Thus, it was verified that the Sugama's modified 
operator works correctly in the Monte Carlo code for multi-ion-species cases. 

\begin{table}[htb]
  \begin{center}
    \caption{Parameters in the 4-species plasma simulation}
    \begin{tabular}{|l||c|c|c|c|} \hline
	& D & T & He & C \\ \hline \hline
Charge $Z_a$ & +1 & +1 & +2 & +6 \\ \hline
Mass  $m_a$ [relative to $H$] & 2   &  3   & 4   &  12 \\ \hline
Density $n_a$ [$10^{19}$m$^{-3}$] &  1.00 & 1.00 & 0.10 & 0.04 \\ \hline
Temperature $T_a$ [keV]  & 2.0  & 2.2  & 1.8  & 1.8 \\ \hline
${\rm Initial}\ \delta n_a/n_a$  & 0.10  & 0.10 & 0.10 & 0.05 \\ \hline
${\rm Initial}\ \delta u_a/v_a$  & 0.10  &-0.10 & 0.10 & -0.20 \\ \hline
${\rm Initial}\ \delta T_a/T_a$  & 0.05  &-0.05 &-0.10 & 0.10 \\ \hline
     \end{tabular}  \label{tbl:2}
  \end{center}
\end{table}

\begin{figure}[tb]
  \includegraphics[width=7cm]{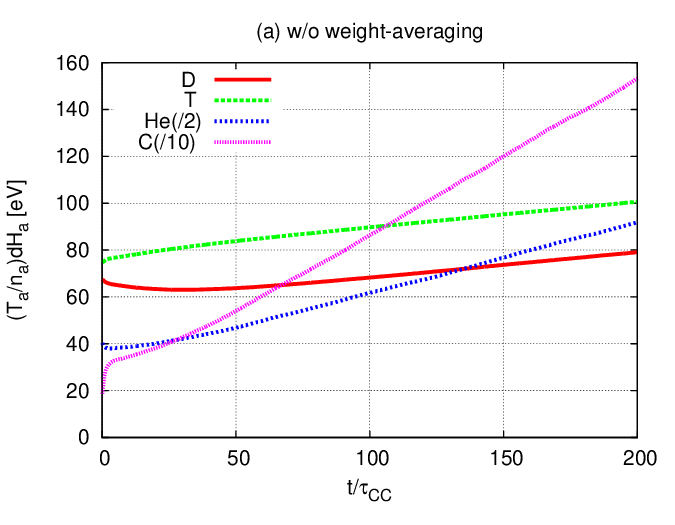}
  \includegraphics[width=7cm]{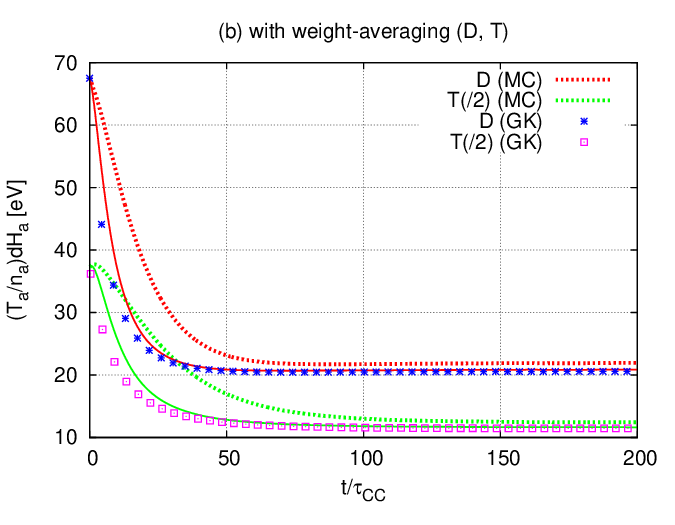}
  \includegraphics[width=7cm]{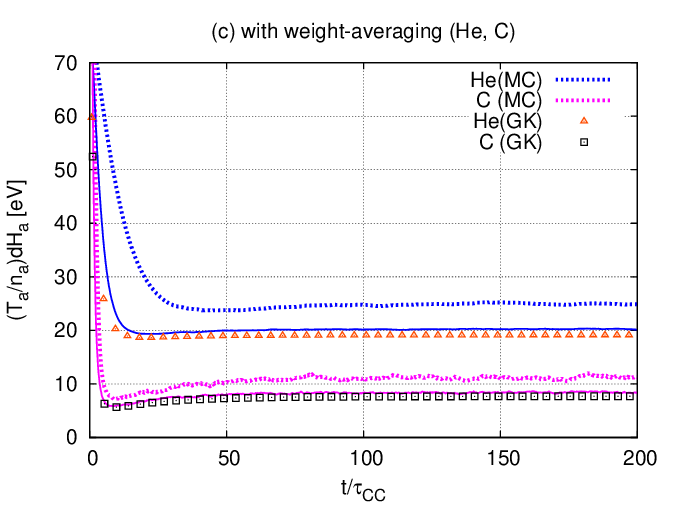}

\vspace{1cm}
  \caption{Time evolution of the quantities $(T_a/n_a)\delta H_a$ defined in Eq. (\ref{h4}) in the 4-species calculation. 
Figure \ref{fig:7}(a) is the case without weight averaging, and \ref{fig:7}(b) and (c) show the cases with averaging, 
$f_W=1$ (thick dotted lines) and $f_W=5$ 
(thin solid lines).  In Figs. \ref{fig:7}(b) and (c), $\delta H_a$ obtained from the continuum gyrokinetic 
$\delf$ code (GK) are also plotted by points.}
  \label{fig:7}
\end{figure}

Second, for the benchmark of the H-theorem in a plasma including 3 or more particle species, 
it is more convenient to observe the time evolution of the following quantities $\delta H_a$ and their sum $H$ 
than directly checking the relation Eq.(\ref{eqbolth}), 
\begin{eqnarray}
\dd{H}{t}&=& \sum_a T_a\sum_b \int d^3v\frac{\delf_a}{f_{Ma}}{C_{ab}}(\delf_a)\nonumber\\
&=& \sum_a \frac{T_a}{2}\dd{}{t} \int d^3v\frac{\delf_a^2}{f_{Ma}}\equiv\sum_a 
\frac{T_a}{2}\dd{}{t}\delta H_a\leq 0 .\label{h4}
\end{eqnarray}
In Figure \ref{fig:7}(a), the time evolution of $(T_a/n_a)\delta H_a$ in the Monte Carlo simulation is plotted. 
Unexpectedly, $\delta H_a$ of all the four species, and therefore their sum $H$, increased monotonously 
after $t>30 \hat{\nu}_{CC}^{-1}$, and the increase continued even after $\delta u$ and $\delta T/T$ 
reached the stationary state, $t>150\hat{\nu}_{CC}^{-1}$. This tendency,
which contradicts to the H-theorem, has been pointed out by Brunner\cite{brun} in the discussion on the  
``weight spreading'' problem caused by the Monte Carlo collision operator for the two-weight $\delf$ method.
Because of the approximation of the test-particle operator $C^{T0}_{ab}$, which is originally the second-order
partial differential equation [Eqs. (3), (4) and (27) in Ref.\cite{sgm-c}], by the Langevin equation-like 
random walk of simulation markers in the velocity space, the weight $w_i$ of each marker tends to 
spread from its ensemble average value, $\overline{W}(\mathbf{v})$, i.e., $w_i(\mathbf{v}_i)=
\overline{W}(\mathbf{v}=\mathbf{v}_i)+\tilde{w}_i$. Note here that the ensemble average value of $\tilde{w}_i$
over all markers is zero. Then, the numerical evaluation of $\delta H_a$ in the $\delf$ Monte Carlo code becomes
\begin{eqnarray}
\delta H_a&=&\int d^3v \frac{\delf_a^2}{f_{Ma}}\simeq \sum_i \frac{w_{a,i}^2}{p_{a,i}}\nonumber\\
&=& \sum_i \frac{\overline{W}(\mathbf{v}_{i})^2}{p_{i}}+\sum_i \frac{\tilde{w}_i^2}{p_i},\label{h5}
\end{eqnarray}
where the subscript $a$ to represent the particle species has been suppressed in the second line.
Note that since the linearized collision operator does not change the weight $p_i$, one should consider 
only the weight spreading of $w$ in the present case. Though the ensemble average part, $\delta \bar{H}_a=
\sum_i \overline{W}(\mathbf{v}_i)^2/p_i$, obeys the H-theorem and $\sum_a \delta \bar{H}_a$ is a 
decreasing function,  the contribution from the $\tilde{w}_i^2$-part increases proportionally to time $t$, 
as Brunner has pointed out.
This apparent breakdown of H-theorem by weight spreading can be restored by the weight averaging 
method which has also been developed by Brunner in the same reference. 
Actually, FORTEC-3D code for single-species plasma has already adopted the weight averaging method\cite{f3d3}.
In the simulation shown in Figs. \ref{fig:6} and \ref{fig:7}(a), the weight averaging has been turned off intentionally  
to see the effect of weight spreading on H-theorem. If the weight averaging is turned on, the marker 
weights are modified as follows:
(the details of the definition of the average weight field $W_a(\mathbf{v})$ is described in \cite{brun,f3d3})
\begin{equation}
w'_i=\gamma_a W_a(\mathbf{v}_i)+(1-\gamma_a)w_i,
\end{equation}
where the damping rate $\gamma_a$ is controlled by a input parameter $f_W$ as follows:
\begin{equation}
\gamma_a\equiv 1-\exp(-f_W\Delta t \hat{\nu}_{aa}),\label{defgam}
\end{equation}
so that the variance of marker weight $\tilde{w}^2$ damps exponentially in the time scale of  $1/(f_W \hat{\nu}_{aa})$.
Larger $f_W$ means stronger averaging on the marker weights. Figures \ref{fig:7}(b) and (c) show the $\delta H_a$ in the
simulations with $f_W=1$ (modest weight averaging case) and $f_W=5$ (strongly averaging case), as well as 
those evaluated in the gyrokinetic simulation. It should be noted here that the gyrokinetic $\delf$ code is free from the 
weight-spreading problem, since it is not a Monte Carlo code. In these figures, one can see that $\delta H_a$
of four species decreased with time by adopting the weight average scheme, 
and they converged to stationary values at $t>150 \hat{\nu}_{CC}^{-1}$, when $\delta u_a$ and $\delta T_a/T_a$
reached to a steady state. Thus the H-theorem, $d{H}/dt=\sum_aT_a(d\delta H_a/dt)\leq 0$,  was restored and
$d(\delta H_a)/dt\rightarrow 0$ as $t\rightarrow\infty$.
As the strength of weight averaging $f_W$ increases, the time evolution and the steady state values of
$\delta H_a$ approaches to those observed in the continuum gyrokinetic simulation. It is then confirmed that the
weight averaging method correctly remove the contribution of  the $\tilde{w}^2$-part to $\delta H_a$ 
in Eq.(\ref{h5}). 

\begin{figure}[tb]
  \includegraphics[width=7cm]{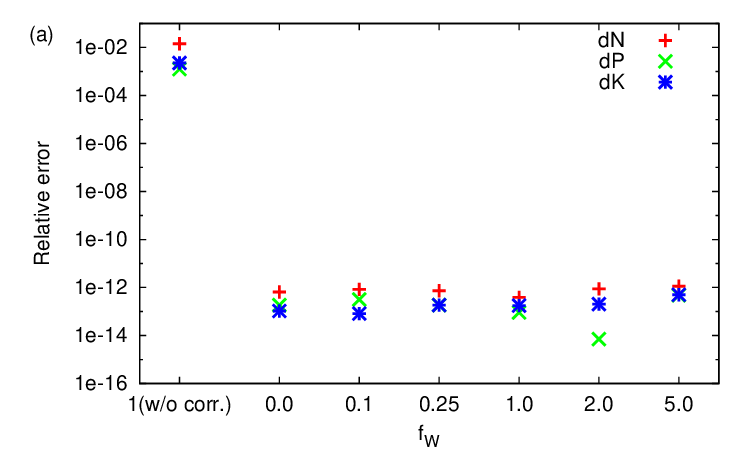}
  \includegraphics[width=7cm]{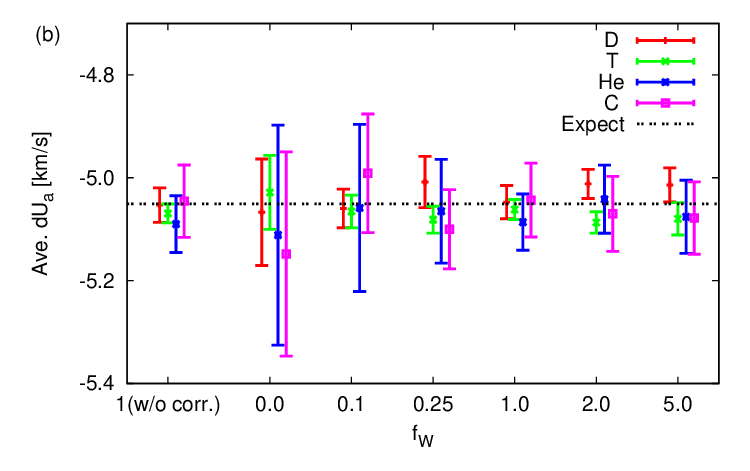}
  \includegraphics[width=7cm]{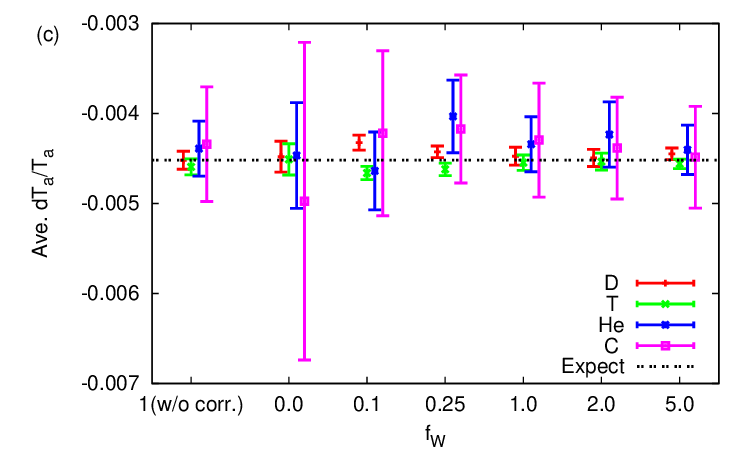}
  \includegraphics[width=7cm]{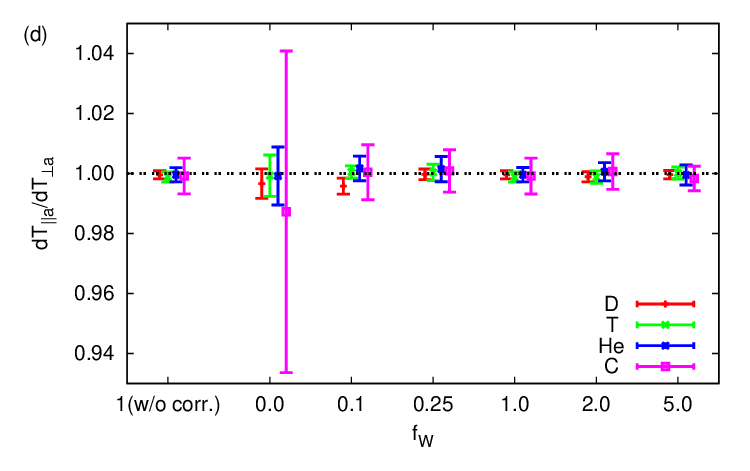}

\vspace{1cm}
  \caption{Dependence on the strength parameter of weight averaging method $f_W$ on the 
time average of (a): relative error in the conserved quantities $\delta N$, $\delta P$ and $\delta K$ (the same as 
the quantities shown in Fig. \ref{fig:3}), (b): mean flow $\delta U_a$, (c): temperature perturbation $\delta T_a/T_a$,
and (d): temperature anisotropy $\delta T_{\parallel a}/\delta T_{\perp a}$, in the 4-species plasma. 
The averages and standard deviations of these quantities were evaluated from 
$t=250$ to $300 \hat{\nu}_{CC}^{-1}$. $f_W=0$ means the no averaging, and $f_W=5$ is the strongest averaging case.
The leftmost points represent the $f_W=1$ case without using the correction scheme in $C_{ab}^F$ for the conservation 
property. Expectation values in Figs. \ref{fig:8}(b) and \ref{fig:8}(c) are estimated according to Eq. (\ref{exp_ut}).}
  \label{fig:8}
\end{figure}

Though it is demonstrated that the weight averaging method works well to recover the H-theorem, 
one may be afraid of possible side effect of the weight averaging method in transport simulation.
Therefore, we checked the dependence of several quantities on the strength of averaging, $f_W$. 
In Figure \ref{fig:8}(a) to (d), time average values of $\delta u_a$ and so on between $t=250$ to $300 
\hat{\nu}_{CC}^{-1}$ are compared among the simulations with $f_W$ varied from 0 (no averaging) to 5. 
Since the weight averaging method is constructed so as to conserve the particle number, momentum, and energy, 
it does not deteriorate the conservation property of $\delta N$, $\delta P$, $\delta K$ 
(the same quantities as in Fig. \ref{fig:3}) 
as shown in Fig. \ref{fig:8}(a). In Figs. \ref{fig:8}(b) and \ref{fig:8}(c), time average of $\delta u_a$ and $\delta T_a/T_a$ 
are compared. Here,
the error bar is estimated from the standard deviation. It is found that if the weight averaging is turned off
or very weak, the average values of $\delta u_a$ and $\delta T_a/T_a$ deviate from the expectation values of them,
and the error bars become larger for heavier ion species (He and C). 
By applying the weight averaging with
enough strength ($f_W\geq 1.0$), the error level in $\delta u$ and $\delta T/T$  reduced, and the mean values converged to their expectations. In Fig. \ref{fig:8}(d) we checked the temperature anisotropy  
\begin{eqnarray*}
\delta T_{\parallel a}/\delta T_{\perp a}=\int d^3v v^2_\parallel \delf_a\big/\int d^3v (v^2_\perp/2) \delf_a.
\end{eqnarray*}
Again, it is found that the error level in $\delta T_{\parallel a}/\delta T_{\perp a}$ is larger for heavier ions but
can be suppressed by the weight averaging. In Figs. \ref{fig:8}(a)-(d), the simulation results of $f_W=1$ without using the 
correction scheme in $C_{ab}^F$ for the conservation property are also shown for comparison. 
As we have found in the 2-species simulations in Section \ref{sec41}, the correction of the field-particle operator 
does not affect the stationary-state average values of $\delta u$ and $\delta T$ but effectively reduces the 
accumulation of error in the quantities $\delta N$, $\delta P$, and $\delta K$. 

To summarize, adopting the weight averaging method 
improve the S/N ratio in evaluating the velocity moments of $\delf$ without  
deteriorating the conservation property of particle numbers, momentum, and energy. It also serves to
restore the H-theorem. Moreover, it was demonstrated that the weight-averaging scheme and the 
correction scheme in the field-particle operator to keep the conservation property can coexist.
However, it should be noted that, in the actual application of the $\delf$ simulation for 
neoclassical transport calculation, too strong averaging results in virtual increase of collision frequency.
In torus plasmas especially in low-collisionality condition, there appears a localized variation of $\delf$ near
the trapped-passing orbit boundary in the velocity space. Strong weight averaging will smooth out the 
large variation of $\delf$ across the boundary, and it will affect the evaluation of the neoclassical fluxes. 
According to the experience of neoclassical transport simulations by single-species FORTEC-3D code,
a practical criterion of the strength of averaging without side effect is $f_W=$0.5-2.0. 

\begin{figure}[tb]
  \includegraphics[width=7cm]{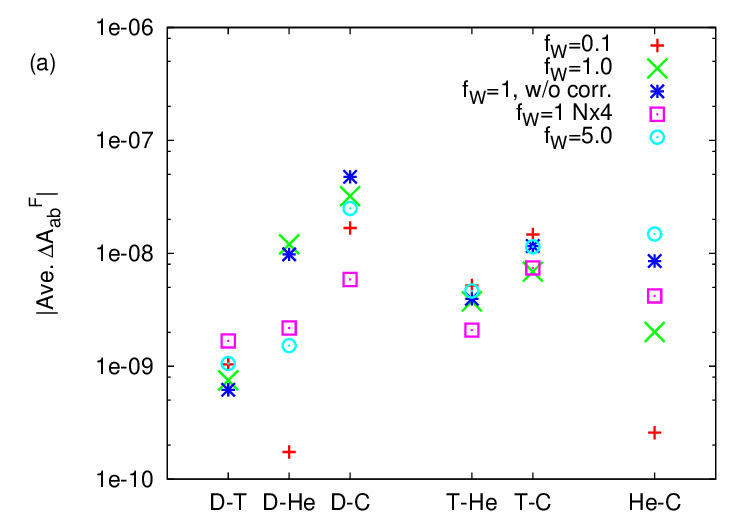}
  \includegraphics[width=7cm]{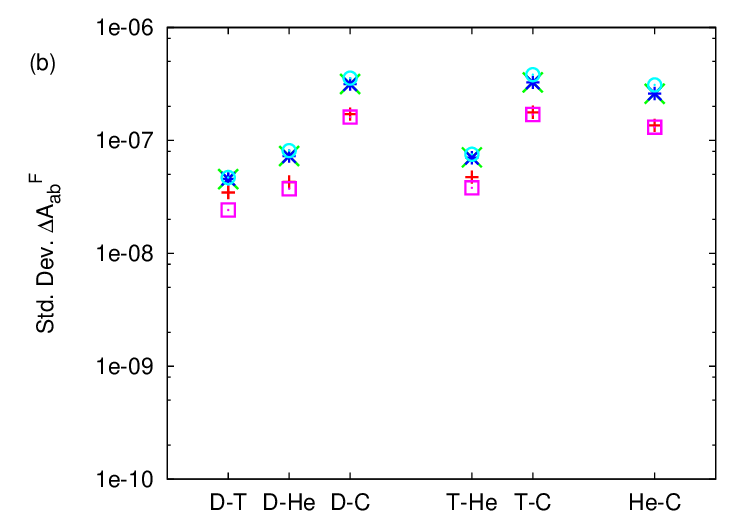}
\vspace{1cm}
  \caption{(a) Average and (b) standard deviation of $\Delta A^F_{ab}$ defined in Eq. (\ref{adj_f}) 
for unlike-species collisions in the 4-species simulations with varying the strength of weight averaging 
($f_W=$0.1, 1, or 5).  The correction method in $C^{F}_{ab}$ was turned on in all the simulations 
except for the ``w/o corr.'' case. The symbols of ``$N\times 4$'' 
represents the result using four times the number of simulation markers than the others. 
The average and standard deviation were evaluated by sampling $\Delta A^F_{ab}$ from the last $50 \hat{\nu}_{CC}^{-1}$
part of each simulation ($2\times 10^5$ samples). } \label{fig:9}
\end{figure}

Finally, the self-adjointness of the linearized collision operator was tested in the following way. 
Since the adjointness property of the test-particle operators $C^{T0}$ and $C^{T}$ is 
difficult to check numerically in the Monte Carlo calculation, the self-adjointness of the 
field-particle operator $C^F$, Eq.(\ref{eqadjf}), was checked. It should be noted that the self-adjoint relations 
of $C^{T0}$ and $C^{T}$ 
are utilized in the derivation of the numerical representations of $C^{T}$ and $C^F$, in Eqs. (\ref{dtpct0}),  
(\ref{eqdv}) and (\ref{eqdw}), respectively. 
If the change of $\delf_a$ by the field-particle part is expressed formally as $(d/dt)|_{C^{F}_{ab}}\delf_a$,
time integral of [LHS of Eq.(\ref{eqadjf}) $-$ RHS of Eq.(\ref{eqadjf})] is numerically evaluated as
\begin{eqnarray}
\Delta A^F_{ab}&\equiv&\int_t^{t+\Delta t}dt \left[
T_a\int d^3v\frac{\delta f_a}{f_{Ma}}\left.\dd{}{t}\right|_{C^{F}_{ab}}\delta f_a -
T_b\int d^3v\frac{\delta f_b}{f_{Mb}}\left.\dd{}{t}\right|_{C^{F}_{ba}}\delta f_b\right]
\nonumber\\
&&\bigg/\left|T_a\int d^3v\frac{\delta f_a^2}{f_{Ma}}\cdot T_b\int d^3v\frac{\delta f_b^2}{f_{Mb}}
\right|^{1/2}\nonumber\\
&\simeq&\left[T_a \int d^3v \left\{\frac{\delta f_a^{(F)2}}{f_{Ma}}-
\frac{\delta f_a^{(T)2}}{f_{Ma}}\right\}-
T_b \int d^3v \left\{\frac{\delta f_b^{(F)2}}{f_{Mb}}-\frac{\delta f_b^{(T)2}}{f_{Mb}}\right\}\right]
\nonumber\\
&&\bigg/\left|2T_a\int d^3v\frac{\delta f_a^{(F)2}}{f_{Ma}}\cdot T_b\int d^3v\frac{\delta f_b^{(F)2}}{f_{Mb}}\right|^{1/2},
\label{adj_f}
\end{eqnarray}
which is normalized in a similar way as $\Delta_{ab}^{(adj)F}$ in Refs. \cite{nnm,nkt}. 
Here, $\delf_a^{(T)}$ and $\delf_a^{(F)}$
represent the distribution function after operating only the $C^T_{ab}$ part and after operating both $C^{T}_{ab}+
C^{F}_{ab}$ parts, respectively. $\Delta A^F_{ab}$ evaluates the 
relative numerical error in the self-adjointness of $C^{F}_{ab}$ per one operation of $C^{F}_{ab}$. 
For like-species collisions, $\Delta A^F_{aa}=0$ by definition.  
Figure \ref{fig:9} shows the average and standard deviation of $\Delta A^F_{ab}$  for unlike-species collisions in the 
4-species plasma simulation. These are the case with the time step size $\Delta t=2.5\times 10^{-4}/\hat{\nu}_{CC}$
and $N_m=4.8\times 10^5$ per a species, except for the ``$N\times 4$'' case in the figures. 
The effect of turning on/off the correction scheme in $C^F_{ab}$ to maintain the conservation 
property was also compared for the case of $f_W=1$. 
It is found that the average and standard deviation of $\Delta A^F_{ab}$ are $O(10^{-9}\sim 10^{-7})$ and 
$O(10^{-8}\sim 10^{-6})$, respectively, and  the amplitude tends to be larger for the combination 
of larger mass ratio such as $(a,b)=$(D,C) and (T,C). 
The adaptation of the strong weight averaging ($f_W\geq 1$) resulted in increasing the standard deviation 
of $\Delta A^F_{ab}$ about factor 2 but the fluctuation still remained small, $<10^{-6}$, 
and it became smaller if  the number of simulation markers was increased. 
It should be pointed out that, though the complete self-adjoint nature of the field-particle operator as Eq. (\ref{eqadjf})
is broken when the correction scheme in $C^F_{ab}$ is adopted, it did not affect in practice on the error level of the 
self-adjointness of the $C^F_{ab}$ operator.
Though the error level of $\Delta A^F_{ab}$ is larger than that of $\delta N$, $\delta P$ and $\delta K$ 
(error in conservation of particle number, momentum, and kinetic energy), the self-adjoint property 
of the field-particle operator is proved to be realized with enough accuracy in the numerical simulation. 

\section{Summary}
In this paper, a numerical method of the linearized Coulomb collision operator for multi-ion-species plasmas
in particle-based $\delta f$ Monte Carlo code have been presented, which satisfies the self-adjointness of the 
operator and the H-theorem even if the particle species have different temperatures. 
In the benchmarks between a full-f PIC code which uses Nanbu-model 
Monte Carlo collision term and a $\delf$ continuum gyrokinetic code which uses the same modified Landau 
operator devised by Sugama, it has been demonstrated that the processes of the damping of the 
mean flow and the thermalization of each particle species towards 
a stationary state agree well with one another. The conservation properties of particle number, momentum, 
and energy of the linearized operator were satisfied with high accuracy, within the rounding error level.
Thus the numerical method to implement Sugama's modified collision operator     
in a $\delf$ Monte Carlo code has been verified. It is found that the weight-spreading phenomenon happening in the $\delf$ code 
deteriorates the  numerical reproducibility of the H-theorem, and this can be restored by adopting the weight-averaging method 
which is already used in the $\delf$ neoclassical transport code, FORTEC-3D. 

The motivation of developing the numerical method of multi-ion-species collision operator is to construct a neoclassical 
transport simulation code for nuclear fusion reactors. To evaluate the neoclassical radial particle and energy fluxes 
as well as bootstrap current in multi-species plasmas, it is expected that the $\delf$ simulation code should solve
the time integral of drift-kinetic equation up to several tens of collision time $\tau_{ab}$, which varies 
according to the combination of particle species $(a,b)$. As the mass ratio of two particles species becomes larger,
the thermalization process becomes slower than the momentum transfer process
and it requires many time steps of simulation to reach a steady-state solution.
Also, the large mass ratio allows that ion species have small but finite different temperature. 
The present collision operator which satisfies the conservation property and the H-theorem forms the reliable basis
for such a long-duration neoclassical transport simulation for multi-species plasmas. 

There are several remaining issues we have not considered here to develop such a multi-species neoclassical transport code.
First, it is required to consider the Coulomb logarithm for unlike-species collisions, $\ln \Lambda_{ab}$.
Throughout this article, we have used the constant value $\ln \Lambda_{ab}=18$ for simplicity, but 
$\ln \Lambda_{ab}$ and therefore the collision time $\tau_{ab}$ vary about $10\sim 20\%$ according to
the analytic models for $\ln \Lambda_{ab}$. We will adopt the analytic model proposed by Honda\cite{honda-ln} 
which is applicable for wide range of plasma parameters. The second problem is the increase in the calculation time
in multi-species collision operator. Since one needs to solve all combinations of the species contained in a plasma, 
the computation cost for $N$-species plasma is about $N^2$ times of that required in a single-species case. Therefore, it is 
preferable that the collision operator calculation is parallelized. Also, the amount of random numbers required 
in $N$-species plasma increases $N^2$ times. We have already implemented the parallel 
pseudo-random number generation scheme of Mersenne Twister\cite{mt-dynam,mt-stk} in our code. Therefore
the increase in the time cost of random number generation can be negligible. The parallelization of the multi-species
collision operator is implemented by MPI and OpenMP hybrid-parallel programming. Further optimization can be realized by
optimizing corrective communications which are required to evaluate integrals such as Eqs. (\ref{dut0}) and (\ref{det0}).

Finally, in this article we have not tested the linearized collision operator for electron-ion case. As it is explained in \ref{apdx_a}, 
the present Monte Carlo test-particle operator $C^{T0}_{ab}$ has a numerical problem when the 
normalized velocity of simulation marker is slow, $x_a=v/v_a\simeq 0$, and the ratio of thermal velocity of two species are large, $\alpha_{ab}=v_a/v_b\gg 1$. 
The main cause of the numerical problem is because the pitch-angle and energy diffusion collision frequency $\nu_{D}^{ab}$ and
$\nu_v^{ab}$ are functions of $v$. If  $\alpha_{ab}\gg 1$, the numerical problem cannot easily be resolved by simply 
reducing the time step size $\Delta t$. A similar problem has also been reported in the continuum gyrokinetic $\delf$  
code which adopts Sugama's modified operator for electron-ion collision case\cite{nkt}. 
Though the present Monte Carlo method works properly  for the 
ion-ion collisions with large mass ratio such as H-Fe or H-W as it was demonstrated, it was found 
that larger correction is required in $C^F_{ab}$ to keep the conservation properties as the mass-ratio becomes larger, 
as shown in Fig. \ref{fig:5}.
To treat the electron-ion collisions in the Monte Carlo neoclassical transport,
we will have to adopt a large-mass-ratio approximation collision operator to avoid the numerical problem at $x_a\simeq 0$.
For single-ion species case, it has been demonstrated that the approximation of $C_{ei}$ in FORTEC-3D
by a large-mass-ratio limit model, which approximates $C_{ei}$ by pitch-angle scattering + friction force between electrons 
and ions with finite ion mean flow, reproduces the correct electron neoclassical transport\cite{hboz}.
The numerical way to Sugama's modified operator for electrons with multiple ion species in a large-mass-ratio limit, 
which is also constructed to satisfy the self-adjoint property, will be reported in another paper.

\section*{Acknowledgment}

\appendix
Part of the simulation was carried out using the HELIOS supercomputer system at International Fusion
Energy Research Centre, Aomori, Japan, under the Broader
Approach collaboration between Euratom and Japan, implemented by Fusion for Energy and JAEA.
It was also carried out using Plasma Simulator in NIFS, under the auspices of the NIFS Collaboration Research 
programs (NIFS14KNTT026, NIFS15KNST079, NIFS18KNST132, NIFS18KNTT045).
This work was supported in part by JSPS Grants-in-Aid for Scientific Research Grant No.19H01879.

\section{Implementation of the test-particle operator $C^{T0}_{ab}$ in the Monte Carlo method}
\label{apdx_a}
\setcounter{figure}{0}

Although the numerical implementation of the test-particle collision operator in Monte Carlo codes appears in many 
articles, here we would like to present explicitly how the random walks of simulation markers 
in the velocity space $(v'^2_i,\xi'_i)=(v^2_i+\Delta v^2_i,\xi_i+\Delta \xi_i)$ are given in the $C^{T0}_{ab}$ operator in 
FORTEC-3D code, for readers' sake.

The original form of  $C^{T0}_{ab}$ is\cite{sgm-c,boz}
\begin{eqnarray}
C^{T0}_{ab}(\delf_a)&=&\nu_D^{ab}(v)\mathcal{L}\delf_a+\mathcal{C}^{ab}_v(\delf_a)\nonumber\\
&=&\frac{\nu_D^{ab}(v)}{2}\dld{}{\xi}\left[(1-\xi^2)\dld{}{\xi}\right]\delf_a\nonumber\\
&&+\frac{1}{v^2}\dld{}{v}\left[v^2\nu_v^{ab}(v)\left(v\delf_a+\frac{v_a^2}{2}\dld{}{v}\delf_a
\right)\right],\label{a1}
\end{eqnarray}
where we use $(v=|\vv|,\xi=\vpr/v)$ as the velocity variables. Note that Eq. (\ref{a1}) is already averaged 
over the gyro-phase. The first and second terms represent the Lorentz
pitch-angle scattering term and the energy diffusion term, respectively, and 
\begin{align}
\nu_D^{ab}(v)&=\hat{\nu}_{ab}\frac{\Phi(x_b)-G(x_b)}{x_a^3},\nonumber\\
\nu_v^{ab}(v)&=\hat{\nu}_{ab}\frac{2G(x_b)}{x_a},\nonumber\\
\Phi(x)&=\frac{2}{\sqrt{\pi}}\int_0^x dt~ \exp(-t^2),\nonumber\\
G(x)&=\frac{\Phi(x)-x\dd{}{x}\Phi(x)}{2x^2},\nonumber\\
\hat{\nu}_{ab}&=n_be_a^2e_b^2\ln\Lambda_{ab}/(4\pi m^2_a\epsilon_0^2v_a^3).\nonumber
\end{align}
Note here that $x_a\equiv v/v_a$ and $x_b\equiv v/v_b$. In a Monte Carlo simulation, the random walk of $(\Delta v^2,
\Delta \xi)$ of each marker velocity is given so that the time change in the average and the variance of 
$(v^2,\xi)$ equal to those expected from Eq. (\ref{a1}). For example, by substituting $\delf_a(\vv=\vv_i)=\delta(v-v_i)
\delta(\xi-\xi_i)$, the expectation values of the changes in $v^2$ and $v^4$ of the markers at $v=v_i$ can be 
estimated as follows:
\begin{align}
\fave{\frac{\Delta v^2}{\Delta t}}&=\int d^3v~ v^2\mathcal{C}_v^{ab}(\delf_a)\nonumber\\
&=-2\int d^3v~ v^2\nu_v^{ab}\delf_a+v_a^2\int d^3v~ \delf_a\left(3\nu_v^{ab}+v\dld{}{v}\nu_v^{ab}
\right)\nonumber\\
&=\nu_v^{ab}(v_i)v^2_i\left(-2+3\frac{v_a^2}{v_i^2}\right)+v_iv_a^2\dld{}{v}\nu_v^{ab}(v_i)\nonumber\\
&=-2\nu_v^{ab}(v_i)v^2_i\left[1-\frac{\alpha_{ab}^2\dd{}{x_b}\Phi(x_b)}{2x_bG(x_b)} \right],\label{a2}\\
\fave{\frac{\Delta v^4}{\Delta t}}&=\int d^3v~ v^4\mathcal{C}_v^{ab}(\delf_a)\nonumber\\
&=\nu_v^{ab}(v_i)v^2_i\left(-4v^2_i+10v^2_a\right)+2v^3_iv^2_a\dld{}{v}\nu_v^{ab}(v_i).\label{a3}
\end{align}
where $\alpha_{ab}=v_a/v_b$.
From these equations, one can estimate the variance of $v^2$ as follows:
\begin{align}
\fave{\frac{\Delta \sigma^2_{v^2}}{\Delta t}}&=\fave{\frac{\Delta v^4}{\Delta t}}-\fave{\frac{(\Delta v^2)^2}{\Delta t}}
=\fave{\frac{\Delta v^4}{\Delta t}}-2v^2\fave{\frac{\Delta v^2}{\Delta t}}\nonumber\\
&=4\nu_v^{ab}(v_i)v^2_iv^2_a.\label{a4}
\end{align}
Then, the $\mathcal{C}^{ab}_v(\delf_a)$ operator is modeled as a Langevin equation of the simulation marker velocity,
\begin{equation}
\Delta v_i^2=-\gamma^{ab}_{v^2}(v_i)v_i^2\Delta t+\delta R\sqrt{D^{ab}_{v^2}(v_i)\Delta t},\label{a5}
\end{equation}
where $\gamma^{ab}_{v^2}(v_i)=-$[Eq. (\ref{a2})]$/v_i^2$ is the friction coefficient, 
$D^{ab}_{v^2}(v_i)=$[Eq. (\ref{a4})] is the diffusion coefficient, 
and $\delta R$ is a random number of which average is zero and the standard deviation is unity.
In our calculation, $\delta R$ is simply $\pm 1$ with equal probability. It should be noted that we treat the 
random walk not in $v$, but in $v^2$ here. The reason is that, if we derive a Langevin equation for $\Delta v$ as 
Eq. (\ref{a5}), then the drag force term $\gamma_v^{ab}v\Delta t$ diverges at $v\rightarrow 0$ and 
is difficult to treat numerically.
One can construct the representation of random walk of $\Delta \xi$ by following the same way as
Eq. (\ref{a2}) - (\ref{a5}). The result is,
\begin{align}
\Delta \xi_i&=-\gamma^{ab}_{\xi}(v_i)\xi_i\Delta t+\delta R\sqrt{D^{ab}_{\xi}(\vv_i)\Delta t},\nonumber\\
\gamma^{ab}_{\xi}(v_i)&=\nu_D^{ab}(v_i),\label{a6}\\
D^{ab}_{\xi}(\vv_i)&=(1-\xi_i^2)\nu_D^{ab}(v_i).\nonumber
\end{align}

In practice, an important property that the Monte Carlo pitch-angle scattering 
operator (\ref{a6}) have is that if $|\xi_i|\leq 1$, then the range of the
map $\xi_i'=C^{T0}_{ab}(\xi_i)$ is also bounded to $|\xi'_i|< 1$. One should note that 
the range of the map is not same as the domain of definition: $\xi_i\in[-1,+1]$, 
but is smaller than the domain, i.e., $\xi_i'\in[-1+\epsilon_\xi,+1-\epsilon_\xi]$, where
$\epsilon_\xi=0.5\nu_D^{ab}\Delta t$\cite{boz}. Therefore,
the Monte Carlo pitch-angle scattering operator is correct only if $\nu_D^{ab}\Delta t\ll 1$. 
However, this condition is not always satisfied in multi-species plasma simulations using a common $\Delta t$ for 
the time integrals of all species. It should be emphasized that the $\Delta t$ for the collision operator 
must be the same value for all particle species, because the present Monte Carlo collision operator involves the 
velocity integrals $\delta u_a^{(T0)}$ and $\delta E_a^{(T0)}$ as Eqs. (\ref{dut0}) and (\ref{det0}) to 
evaluate the change in the momentum and energy by $C_{ab}^{T0}$,
and also because the a-b and b-a collisions should be treated simultaneously in the field-particle operator to satisfy the 
momentum and energy conservation. In our Monte Carlo code, for example, chooses 
$\Delta t\sim 10^{-4}\tau_{min}$ where $\tau_{min}$ is the minimum value of $\hat{\nu}_{ab}^{-1}$ from all the 
combinations of $(a,b)$. No matter how small $\Delta t$ is chosen, there is a finite probability that 
$\nu^{ab}_D(v_i)\Delta t\sim 1$, because the simulation markers have a velocity distribution
close to Maxwellian, and $\nu^{ab}_D(v\rightarrow 0)\simeq 0.752\hat{\nu}_{ab}\alpha_{ab}x_a^{-2}$. 
However, it is not efficient to reduce the time step size for all species only for the small fraction of markers which 
violates the condition $\nu_D^{ab}\Delta t\ll 1$. Instead, our strategy is that if $\nu_D^{ab}(v_i)\Delta t> 1$ 
for a simulation marker, then the test-particle operator $C_{ab}^{T0}$ gives a random value of $\xi'_i\in[-1,+1]$,
to mimic a large-angle pitch angle scattering. If $\Delta t$ is chosen small enough, this treatment does not affect the 
simulation result. For example, in the 4-species plasma simulation shown in \ref{sec42}, where we choose 
$\Delta t=2.5\times 10^{-4}\tau_{min}$, only $10^{-4}\%$ of total test-particle collisions met the 
criterion $\nu_D^{ab}\Delta t>1$.

\begin{figure}[tb]
  \includegraphics[width=7cm]{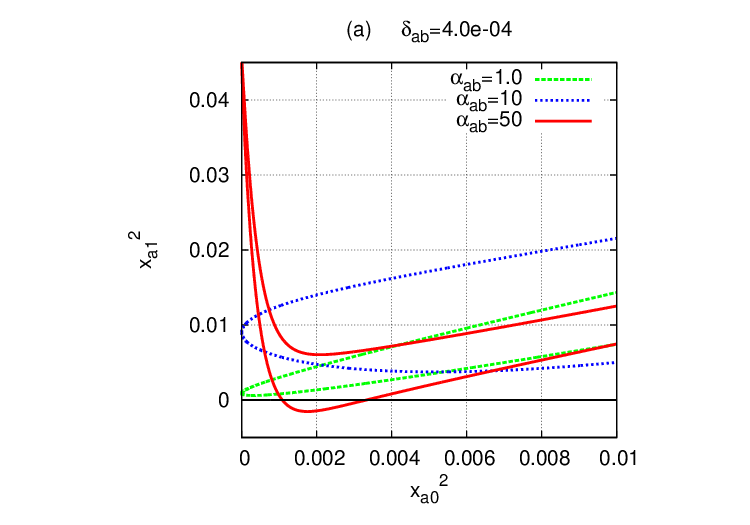}
  \includegraphics[width=7cm]{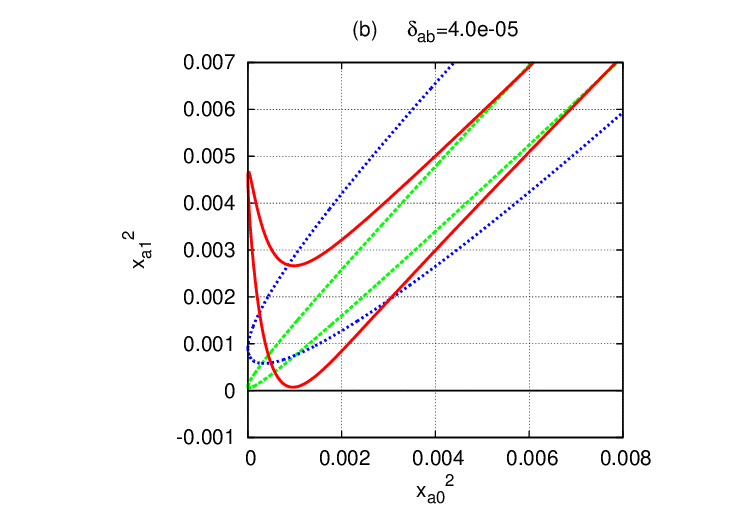}
  \includegraphics[width=7cm]{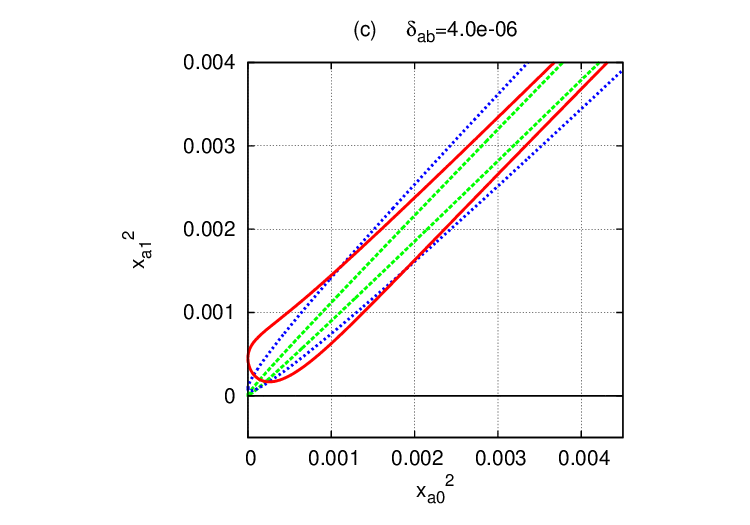}
\vspace{1cm}
  \caption{Profiles of $x_{a1}^2=v'^2_i/v_a^2$ as  functions of $x_{a0}^2=v_i^2/v_a^2$ which are calculated from the 
Monte Carlo test-particle operator $C_v^{ab}$, Eq. (\ref{a5}), for several choices of the parameters $(\alpha_{ab},\delta_{ab})$. 
The upper and lower 
curves for each $\alpha_{ab}$ represent the values of $x_{a1}^2$ when the random number $\delta R=+1$ or $-1$, respectively.
 The region in which $x^2_{a1}<0$ for the $\alpha_{ab}=50$ case in Fig. \ref{fig:A1}(a) means the Monte 
 Carlo operator is incorrect there.}
  \label{fig:A1}
\end{figure}

For the energy diffusion term $C_v^{ab}$, the problem at $v\rightarrow 0$ is more complicated than the Lorentz 
operator, because $\nu_v^{ab}$ is a function of $v$ in itself. 
Considering the Taylor expansion of $\nu_v^{ab}(v)$ at $v\simeq 0$, one finds that 
\begin{align}
\nu_v^{ab}(v\simeq 0)&\simeq \frac{4\hat{\nu}_{ab}\alpha_{ab}}{3\sqrt{\pi}},\label{a7}\\
v'^2_i(v_i\simeq 0)&\simeq v_i^2 \left[1-\frac{8\alpha_{ab}\delta_{ab}}{3\sqrt{\pi}}\right] +
v_a^2\frac{4\alpha_{ab}\delta_{ab}}{\sqrt{\pi}}+4v_av_i\delta R\sqrt{\frac{\alpha_{ab}\delta_{ab}}{3\sqrt{\pi}}},
\label{a8}
\end{align}
where $\delta_{ab}\equiv \Delta t\hat{\nu}_{ab}$. It should be noted that the approximation above is valid
only if $\alpha_{ab}\delta_{ab}\ll 1$. From Eq. (\ref{a8}) one notices that 
\begin{equation}
v'^2_i(v_i\rightarrow 0)=v^2_a\frac{4\alpha_{ab}\delta_{ab}}{\sqrt{\pi}},\label{a9}
\end{equation}
which means that, if $\alpha_{ab}\delta_{ab}$ is not so small, a simulation marker $v_i\simeq 0$ receives an intense 
drag force $-\gamma^{ab}_{v^2}v_i^2\Delta t$. Another problem of the $C_v^{ab}$ operator 
at $v_i\sim 0$ is that the range of $v'^2_i$. Taking the derivative $dv'^2_i/dv_i$  of Eq. (\ref{a8}) with $\delta R=-1$, 
one finds that the minimum value of $v'^2_i$ occurs when 
\begin{equation}
v_i=v_{ip}\equiv\frac{2v_a\sqrt{\frac{\alpha_{ab}\delta_{ab}}{3\sqrt{\pi}}}}{1-\frac{8\alpha_{ab}\delta_{ab}}{3\sqrt{\pi}}}.
\label{a10}
\end{equation}
Though Eqs. (\ref{a8}) and (\ref{a10}) are valid only if $\alpha_{ab}\delta_{ab}\ll 1$, these approximated expressions
suggest that the range of $v'^2_i$ is not bounded to $(0,+\infty)$, but $v'^2_i$ can even become negative
around $v_i=v_{ip}$ in some cases, depending on the parameter $\alpha_{ab}\delta_{ab}$.  Figures \ref{fig:A1}(a-c) show the 
profiles of $x_{a1}^2=v'^2_i/v_a^2$ as  functions of $x_{a0}^2=v_i^2/v_a^2$ which are calculated from Eq. (\ref{a5}) 
for several choices of $(\alpha_{ab},\delta_{ab})$.
It is found that the lower curve of $x^2_{a1}$, which correspond to the cases of $\delta R=-1$, becomes negative 
if $\alpha_{ab}\delta_{ab}>\sim 0.01$. Also, at $x_{a0}^2\rightarrow 0$, it can be seen that $x_{a1}^2$ follows Eq. (\ref{a9}).
In summary, the Monte Carlo implementation of $C_{ab}^v$ by Eq. (\ref{a5}) is not valid around $x_a\simeq 0$ if
$\alpha_{ab}\gg 1$ even if a small time step size $\delta_{ab}=\Delta t\hat{\nu}_{ab}$ is chosen. 
This imposes a more severe condition on the choice of $\Delta t$ than that is required in the Lorentz operator.
Since $\alpha_{ab}=v_a/v_b\sim\sqrt{m_b/m_a}$, this becomes a problem for a light species test-particle collisions with 
very heavy species. In the ion-ion collisions such as H-Fe and H-W in Sec. \ref{sec41}, 
$\alpha_{ab}$ is about 10 at most, and therefore this problem did not matter in practice. For electron-ion collisions, however,
$\alpha_{ab}\sim 50$ and one should choose very small $\delta_{ab}$ to avoid this problem. The other way to avoid the 
problem is to adopt the large-mass-ratio approximation, i.e., the light-to-heavy species test-particle operator is approximated
only by the Lorentz pitch-angle scattering operator and therefore the energy transfer from a heavy to a light species is neglected.
Sugama has shown that there is a way to construct a linearized collision operator in the large-mass-ratio approximation  
which satisfies the particle number and momentum conservation and self-adjointness 
even if $T_a\neq T_b$ \cite{sgm-c}. In the proposed model, although the energy transfer rate from the heavy to light 
species is the same as the original Landau operator, the energy transfer to the opposite direction is zero because
of the large-mass-ratio approximation.

\section{Derivation of Eqs. (\ref{dtpct0})-(\ref{dtpct0p})}\label{apdx_b}
The original form of the projection-part in the test-particle operator $C^T_{ab}$ in Ref.\cite{sgm-c} is given as follows:
\begin{eqnarray}
\mathcal{P}_aC^{T0}_{ab}\delf_a&=&f_{Ma}\left[\frac{m_a\vpr}{T_a}\frac{1}{n_a}\int d^3v'\frac{\delf_a}{f_{Ma}}C^{T0}_{ab}(\vpr' f_{Ma})
\right. \nonumber\\
&&\left. +\left(x_a^2-\frac{3}{2}\right)\frac{1}{n_a}\int d^3v'\frac{\delf_a}{f_{Ma}}\frac{2}{3}C^{T0}_{ab}(x_a'^2f_{Ma}) \right],
\label{pct0_org}\\
C^{T0}_{ab}\mathcal{P}_a\delf_a&=&\frac{m_a\delta u_a[\delf_a]}{T_a}C^{T0}_{ab}(\vpr f_{Ma})
+\frac{\delta T_a[\delf_a]}{T_a}C^{T0}_{ab}(x^2_af_{Ma}),\label{ct0p_org}\\
\mathcal{P}_aC^{T0}_{ab}\mathcal{P}_a\delf_a&=&f_{Ma}\left[\frac{m_a}{T_a}\vpr\delta u_a[\delf_a]\frac{1}{n_a}\int d^3v'
\frac{m_a\vpr'}{3T_a}C^{T0}_{ab}(\vpr' f_{Ma})\right. \nonumber\\
&&\left. +\frac{\delta T_a[\delf_a]}{T_a}\left(x_a^2-\frac{3}{2}\right)\frac{1}{n_a}\int d^3v'\frac{2x_a'^2}{3}C^{T0}_{ab}(x_a'^2f_{Ma}) \right],
\label{pct0p_org}
\end{eqnarray}
where only the parallel component of velocity and mean flow, $\vpr$ and $\delta u_a$, are considered since the 
gyrophase dependence of distribution function $\delf_a$ is neglected in drift-kinetic equation.
We will omit the argument $[\delf_a]$ in $\delta u_a$ and $\delta T_a$ hereafter.

From the self-adjoint property of $C^{T0}_{ab}$ in Eq. (\ref{eqadjt}), one finds 
\begin{eqnarray*}
\frac{1}{n_a}\int d^3v\frac{\delf_a}{f_{Ma}}C^{T0}_{ab}(\vpr f_{Ma})&=&\frac{1}{n_a}\int d^3v \vpr C^{T0}_{ab}(\delf_{a})=
\ddct\delta u_a,\\
\frac{1}{n_a}\int d^3v\frac{\delf_a}{f_{Ma}}\frac{2}{3}C^{T0}_{ab}(x_a^2f_{Ma})&=&
\frac{2}{3n_av_a^2}\int d^3v v^2 C^{T0}_{ab}(\delf_a)=\frac{2}{3v_a^2}\ddct\delta E_a,
\end{eqnarray*}
where $\ddct F$ denotes the time change rate of a quantity F by $C^{T0}_{ab}$.
Substituting these equations to Eq. (\ref{pct0_org}) yields
\begin{equation}
\mathcal{P}_aC^{T0}_{ab}\delf_a=f_{Ma}\left[\frac{m_a\vpr}{T_a}\ddct\delta u_{a}
+\frac{2}{3v_a^2}\left(x_a^2-\frac{3}{2}\right)\ddct\delta E_{a}\right].\label{pcdf}
\end{equation}
Then, integrating Eq. (\ref{pcdf}) over a short time step $\Delta t$ and using the notation (\ref{duabt0}) and (\ref{deabt0}), that is,
\begin{eqnarray*}
\int^{\Delta t}dt~ \ddct\delta u_{a}=\Delta u_{ab}^{(T0)},&&
\int^{\Delta t}dt~ \ddct\delta E_{a}=\Delta E_{ab}^{(T0)},
\end{eqnarray*}
one obtains Eq. (\ref{dtpct0}).

Next, the velocity moments of $C^{T0}_{ab}(\{\vpr,\ v^2\}f_{Ma})$ appeared in Eq. (\ref{pct0p_org}) can be integrated as follows:
\begin{align}
\frac{1}{n_a}\int d^3v\frac{m_a\vpr}{3T_a}C^{T0}_{ab}(\vpr f_{Ma})&=-\frac{4\hat{\nu}_{ab}\alpha_{ab}}{3\sqrt{\pi(1+\alpha_{ab}^2)}},\label{alp1m}\\
\frac{1}{n_a}\int d^3v\frac{2x^2_a}{3}C^{T0}_{ab}(x_a^2f_{Ma}) &=-\frac{8\hat{\nu}_{ab}\alpha_{ab}}{3\sqrt{\pi}(1+\alpha_{ab}^2)^{3/2}},\label{alp2m}
\end{align}
where $\alpha_{ab}$ and $\hat{\nu}_{ab}$ are defined as in Eqs. (\ref{alpab}) and (\ref{nuab}).
$\delta T_a$ in Eqs. (\ref{ct0p_org}) and (\ref{pct0p_org}) is rewritten as follows:
\begin{equation*}
\frac{\delta T_a}{T_a}=\frac{1}{n_a}\int d^3v \left(\frac{2v^2}{3v_a^2}-1\right) \delf_a=\frac{2}{3v_a^2}\delta E_a -\frac{\delta n_a}{n_a}.
\end{equation*}
Then, time integrals of the projections $C^{T0}_{ab}\mathcal{P}_a$ and $\mathcal{P}_aC^{T0}_{ab}\mathcal{P}_a$ over a short 
step size $\Delta t$ is approximated by the trapezoidal rule on $\delta u_a$ and $\delta E_a$ as follows:
\begin{eqnarray}
\int^{\Delta t} dt \delta\{u_a, E_a\} \approx \frac{\delta\{u_a, E_a\}^{(T0)}+\delta\{u_a, E_a\}^{(0)}}{2}\Delta t.
\label{intdelue}
\end{eqnarray}
Finally, by substituting Eqs. (\ref{alp1m}) - (\ref{intdelue}) to (\ref{ct0p_org}) and (\ref{pct0p_org}), one obtains Eqs. (\ref{dtct0p}) and 
(\ref{dtpct0p}). Note that we have used the fact that $\delta n_a$ is unchanged by $C^{T0}$, or  $\delta n_a^{(0)}=\delta n_a^{(T0)}$.

The functions $C^{T0}_{ab}(\vpr f_{Ma})$ and $C^{T0}_{ab}(x^2_a f_{Ma})$ in Eq. (\ref{dtct0p}) can be written down in 
terms of $G(x)$ and $\Phi(x)$ appeared in \ref{apdx_a} as follows: 
\begin{align}
C^{T0}_{ab}(\vpr f_{Ma}) &= -2\hat{\nu}_{ab} (1+\alpha_{ab}^2) \vpr\frac{G(x_b)}{x_a}f_{Ma}(v),\\
C^{T0}_{ab}(x^2_a f_{Ma})&= -\frac{2\hat{\nu}_{ab}}{\alpha_{ab}^2x_a}\left[\Phi(x_b)-x_b(1+\alpha^2_{ab})\dd{\Phi(x_b)}{x_b}
\right]f_{Ma}(v).
\end{align}
Therefore, the time integral of three projection terms in the test-particle operator $C^T_{ab}(\delf_a)$, i.e. 
$\mathcal{P}_aC^{T0}_{ab}\delf_a$, $C^{T0}_{ab}\mathcal{P}_a\delf_a$, and $\mathcal{P}_aC^{T0}_{ab}\mathcal{P}_a\delf_a$, are all proportional to Maxwellian $f_{Ma}$. It has also been shown here that they can be evaluated simply though the 
$\{1,~ \vpr,~ v^2\}$-moments of $\delf_a$ before and after operating $C^{T0}$.
These facts are utilized to represent the time evolution distribution function $\delf_a$ and that of simulation marker weight $w_i$ 
as in Eqs. (\ref{dfdtct}) and (\ref{dwdtct}).

\section{Collision operator in the case of $T_a=T_b$}\label{apdx_c}
It will be instructive to show the form of test- and field-particle collision operators in the case of
$T_a=T_b$, which is frequently assumed in the neoclassical simulations for multi-ion species plasmas.
The same temperature limit corresponds to $\alpha_{ab}=v_a/v_b=\sqrt{m_b/m_a}$ and $\theta_{ab}=1$. 
Therefore, all the terms which are proportional to $(\theta_{ab}-1)$ in $C^T_{ab}$ and  $C^F_{ab}$ drop
in this case. It means that the test-particle operator becomes only the $C^{T0}_{ab}$ 
part given in Eq. (\ref{a1}) and is implemented by the Monte Carlo scheme as explained in \ref{apdx_a}.
Since $\theta_{ab}-1=0$, the source term $S^T_{ab}$ in Eq.(\ref{dwdtct}) becomes
zero and the marker weight $w$ is unchanged in the test-particle part.
Concerning the field-particle part,  the functions $R_{ab}$ and $Q_{ab}$ appeared in Eq. (\ref{eqcf}) 
are also simplified in the $\theta_{ab}-1\rightarrow 0$ limit. Substituting
$(1+\alpha_{ab}^2)=(1+m_b/m_a)$, they become
\begin{align}
R_{ab}(v,\vpr)&=\frac{3\sqrt{\pi}\left(1+\frac{m_b}{m_a}\right)^{3/2}G(x_b)}{n_ax_a}\sqrt{\frac{m_a}{m_b}}\left(\frac{\vpr}{v_a}\right),\\
Q_{ab}(v)&=\frac{\sqrt{\pi}\left(1+\frac{m_b}{m_a}\right)^{3/2}}{2n_ax_b}\frac{m_a}{m_b}
\left[\Phi(x_b)-x_b\Phi'(x_b)\left(1+\frac{m_b}{m_a}\right)\right].
\end{align}
In the $T_a=T_b$ case, the momentum and energy change caused by the test-particle operator
$\delta V^{T}_{ba}$ and $\delta W^{T}_{ba}$ [Eqs. (\ref{eqdv}), (\ref{eqdw})] are evaluated only
from $C^{T0}_{ab}$. Therefore, $\Delta u^{(T)}_{ba}$ and $\Delta E^{(T)}_{ba}$ in Eqs. (\ref{eqdv}) and 
(\ref{eqdw}) are replaced with $\Delta u^{(T0)}_{ba}$ and $\Delta E^{(T0)}_{ba}$, respectively.
In Ref.\cite{koles}, they use $y_b=x_b^2$ and $\phi(y_b)=(2/\sqrt{\pi})\int^{y_b}_0dt e^{-t}\sqrt{t}
=2x_b^2G(x_b)$ to represent the field-particle operator. By noting the numerical factor difference 
in the definitions of $R_{ab}$ and $Q_{ab}$ in the present paper and those in Ref.\cite{koles}, 
it is found that the field-particle operator $C^F_{ab}$ in the $T_a=T_b$ limit is identical to 
that in Ref.\cite{koles}.



\end{document}